\documentclass[aps,prb,reprint,groupedaddress]{revtex4-2}

\usepackage{amsmath}
\usepackage{amsthm}
\usepackage{amssymb}
\usepackage{graphicx}
\usepackage{breqn}
\usepackage{xcolor}

\newcommand{\avg}[1]{\langle #1 \rangle}

\makeatletter
\let\cat@comma@active\@empty
\makeatother

\begin{document}

\title{Single impurity Anderson model out of equilibrium: A two-particle semi-analytic approach}
\author{Jiawei Yan}
\email{yan@fzu.cz}
\affiliation{Institute of Physics of the Czech Academy of Sciences, Na Slovance 1999/2, 182 00 Prague 8, Czech Republic}
\author{V\'aclav Jani\ifmmode \check{s}\else \v{s}\fi}
\affiliation{Institute of Physics of the Czech Academy of Sciences, Na Slovance 1999/2, 182 00 Prague 8, Czech Republic}

\date{\today}


\begin{abstract}
We apply a two-particle semi-analytic approach to a single Anderson impurity attached to two biased metallic leads. The theory is based on reduced parquet equations justified in critical regions of singularities in the Bethe-Salpeter equations. It offers a way to treat one-particle and two-particle thermodynamic and spectral quantities on the same footing. The two-particle vertices are appropriately renormalized so that spurious transitions into the magnetic state of the weak-coupling approximations are suppressed. The unphysical hysteresis loop in the current-voltage characteristics is thereby eliminated. Furthermore, in the linear response regime, we qualitatively reproduce the three transport regimes with the increasing temperature: from the Kondo resonant tunneling through the Coulomb blockade regime up to a sequential tunneling regime. Far from equilibrium, we find that the bias plays a similar role as the temperature in destroying the Kondo resonant peak when the corresponding energy scale is comparable with the Kondo temperature. Besides that, the applied voltage in low bias is shown to develop spectral peaks around the lead chemical potentials as observed in previous theoretical and experimental studies.
\end{abstract}

\maketitle

\section{Introduction\label{sec: introduction}}

Finding the full solutions of the single-impurity Anderson model (SIAM) of a magnetic impurity coupled to a metallic reservoir has been an enduring problem in condensed matter physics since the model was first proposed by P.W. Anderson \cite{PhysRev.124.41}.
The physics behind this model involves competition between formation of the local magnetic moment in strong coupling and local Fermi liquid in weak coupling at low temperatures. The boundary between the two regimes is characterized by an exponentially small Kondo temperature $T_K$ \cite{hewson1997kondo,doi:10.1063/9.0000019}.

Despite the simple form of the Anderson Hamiltonian the solution of the SIAM plays a crucial role in the development of modern theoretical methods for strongly correlated systems since (i) it is one of the few many-body quantum problems that is exactly solvable under some specific conditions \cite{doi:10.1080/00018738300101581} and (ii) it offers an impurity solver for the Dynamical Mean-Field Theory (DMFT) of the Hubbard model \cite{RevModPhys.68.13}.
The  SIAM then serves as a benchmark on which many-body computational methods  are tested. The dynamical mean-field theory built upon the SIAM combined with the density functional theory (DFT) has become one of the most effective ways to implement strong electron correlations into realistic calculations of the electronic structure of solids \cite{RevModPhys.78.865,*doi:10.1063/1.1712502}.
Due to wide applicability of the SIAM, generalizations of the standard SIAM have been proposed for modeling quantum dots in different conditions such as,  (i) impurity with s-wave superconducting leads, representing $0$-$\pi$ junctions, \cite{Meden_2019,*JOSEPHSON1962251,*Zonda2015}, (ii)  complex of quantum dots \cite{PhysRevLett.74.4702,*Fujisawa932},  (iii) impurity  with a spin-orbit coupling \cite{PhysRevB.97.064519}, and (iv) impurity with an external biased voltage to drive the system out of equilibrium \cite{PhysRevLett.70.2601,PhysRevB.78.235110,PhysRevB.81.165115}.
All these  impurity problems cannot be solved exactly  due to the on-site Coulomb repulsion.

Advanced numerical methods have been employed to solve impurity models and to reach trustworthy quantitative results in various situations. For example, quantum Monte Carlo (QMC) \cite{RevModPhys.83.349,gubernatis2016quantum,PhysRevLett.99.236808} provides us with reliable thermodynamic quantities at non-zero temperatures. It is not, however, suitable for accessing real-frequencies and low-temperature spectroscopic properties. It is then replaced there by the numerical renormalization group (NRG) \cite{RevModPhys.47.773,RevModPhys.80.395,PhysRevB.79.085106}  reproducing well the exact low-energy excitations.

It is generally demanding to reach numerically exact solutions even in simple impurity models.  That is why  analytic and semi-analytic approaches have also been widely used to assess the low-temperature behavior of impurity models.   Many-body Green functions proved to become the most suitable tools to achieve this goal.  They may be treated and approximated in various ways.  The equation of motion (EOM) scheme truncates  the hierarchy of equations for many-body Green function at a certain particle  level \cite{PhysRevB.59.9710,PhysRevB.81.165115}.  The standard many-body perturbation theory (MBPT) in the interaction strength  with Feynman is mostly cut  in weak coupling at second order \cite{bruus2004many,haug2008quantum,stefanucci2013nonequilibrium}. Extensions to intermediate coupling can be achieved by  the
GW scheme \cite{HEDIN19701,PhysRevB.77.115333,PhysRevB.79.155110},  the fluctuation-exchange (FLEX) scheme \cite{Bickers:1989ab,*Bickers:1991aa,*Flex1991Chen}  or the parquet approach \cite{Bickers:1991ab,*Bickers:1992aa}. They all sum self-consistently infinite series  of classes of Feynman diagrams. 
None of the methods is, however, applicable in all situations. Hence, new approximate schemes  have been introduced in many-body models constantly \cite{PhysRevX.7.031013,PhysRevLett.101.066804,PhysRevB.86.155130,PhysRevB.92.125145,PhysRevB.75.045324,PhysRevResearch.2.043052,PhysRevB.96.085107}.

The parquet approach singles out from the other ones in that it introduces a two-particle self-consistency in which vertex functions are determined from non-linear (self-consistent) equations.
This feature is extremely important for suppressing unphysical and spurious phase transitions of the weak-coupling constructions.
An obstacle in the wide-spread application of the parquet construction is its complexity and a large number of dynamical degrees of freedom it introduces.
Recently, schemes  to reach numerical solutions of the parquet equations were proposed by using the Matsubara formalism \cite{Li:2019aa,*Li:2016aa,Rohringer:2018aa,Yang:2009aa}.
Numerical results showed quantitative improvements in weak and intermediate couplings but still essentially failed in the strong coupling regime and deep in the critical region of the quantum phase transitions.
One of the present authors introduced the so-called reduced parquet equations to overcome this deficiency \cite{Janis:2007aa,Janis:2008ab,Janis:2017aa,Janis:2017ab,Janis:2019aa,Janis:2020aa}. The reduced parquet equations interpolate reliably between weak and strong couplings of impurity and extended lattice models.  
They capture the most relevant contributions in the critical region of singularities in the Bethe-Salpeter equations.   The long-range fluctuations in the divergent two-particle vertex are treated dynamically in this scheme while its irrelevant short-range fluctuations are replaced by their averages.
As a result, the formalism is significantly  simplified with results showing qualitative agreement with more advanced  numerical methods in the whole parameter space \cite{Janis:2019aa}.

Motivated by these facts we extend here the reduced parquet equations to systems out of equilibrium that will allow us to study transport properties available experimentally \cite{Cronenwett540,cite-key,vanderWiel2105}.
%
%
The reduced parquet equations in their static approximation introduce a renormalization of the bare interaction suppressing the spurious transition to a magnetically ordered state, the magnetic susceptibility remains positive and  the solution is freed of the unphysical hysteresis loop in the current-voltage characteristics of the weak-coupling approximations without a two-particle self-consistency  for out-of-equilibrium systems.
Additionally, the three-peak structure of the equilibrium spectral function is maintained with the correct logarithmic Kondo scaling of the width of the central quasiparticle peak with the interaction strength. We further reveal three transport stages with the increasing of temperature in the linear response regime of this approximation. They are the Kondo resonant tunneling, the Coulomb-blockade regime, and a sequential tunneling \cite{Pustilnik_2004,bruus2004many}.  Farther from equilibrium, we find that the biased voltage inducing the non-equilibrium quantum statistics plays a similar role as temperature in equilibrium in that it destroys the Kondo peak when its value is comparable with the Kondo temperature $T_K$. Besides that, the bias also develops a local spectral peak around the chemical potential of each lead \cite{Cronenwett540,PhysRevLett.70.2601}. These local peaks vanish quickly with the increase of the voltage and finally become unrecognizable.


The  paper is organized as follows. We introduce the model Hamiltonian as well as the Keldysh-Schwinger  non-equilibrium Green function (NEGF) perturbation theory in section \ref{sec: model}.
 The two-particle approach with the reduced parquet equations  formulated in the Keldysh space are introduced in section \ref{sec: 2PSCF}. The real-time formulation to study the steady-state quantum transport problem is presented in section \ref{sec: transport}. Results are discussed  in section \ref{sec: numerics} following by the  concluding section \ref{sec: conclusions}. Important  technical details of the derivations are moved to Appendices.



\section{Model Hamiltonian and NEGF Theory\label{sec: model}}

\subsection{Generic model Hamiltonian\label{sec:generic}}

\begin{figure}
\includegraphics[width=0.8\linewidth]{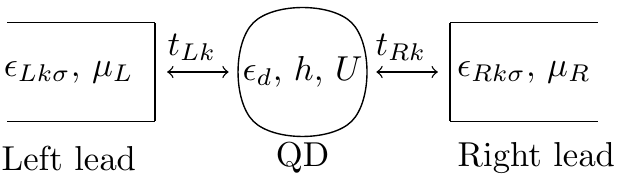}
\caption{Sketch of the transport model, whose Hamiltonian is given by Eq.\eqref{eq: Hamiltonian of transport model}.
\label{fig: model}}
\end{figure}

We describe a single quantum dot (QD) attached to two biased semi-infinite metallic leads, as shown in Fig.\ref{eq: Hamiltonian of transport model}, by the following Hamiltonian 
\begin{equation}\label{eq: Hamiltonian of transport model}
H = \sum_s \left( H^{ld}_s + H_s^{hyb}\right) + H^{dot}, \quad s \in \{L,R\}
\end{equation}
where $H^{ld}_s$, $H^{dot}$ and $H^{hyb}_s$ correspond respectively to the Hamiltonian of the $s$-lead, the QD and the hybridisation between them. The explicit forms of the partial Hamiltonians are
\begin{align}
H^{ld}_s &= \sum_{k\sigma} (\epsilon_{sk\sigma} - \mu_s) c^\dag_{sk\sigma} c_{sk\sigma},\label{eq: Hamiltonian of lead}\\
H^{hyb}_{s} &= -\sum_{k\sigma} \left( t_{sk} d^\dag_\sigma c_{sk\sigma} + t^*_{sk} c^\dag_{sk\sigma} d_\sigma  \right),\label{eq: Hamiltonian of hybridization}\\
H^{dot} &= \sum_\sigma (\epsilon_d - \sigma h)d_\sigma^\dag d_\sigma + U d_\uparrow^\dag d_\uparrow d_\downarrow^\dag d_\downarrow.\label{eq: Hamiltonian of quantum dot}
\end{align}
with $c^{(\dag)}_{sk\sigma}$ and $d^{(\dag)}_\sigma$ the annihilation (creation) operators of the lead and QD electrons, respectively. We denoted $\epsilon_{sk\sigma}$  the dispersion relation of the lead electrons,  $\mu_s$ is  the chemical potential of the $s$-lead, and $\epsilon_d$ is the atomic level of the QD.  The Zeeman magnetic field $h$ splits the spin orientation  $\sigma=\pm 1$ corresponding to up/down spin projection. The hybridization between the QD and the $s$-lead  is $t_{sk}$  and $U$ is the charging energy on the QD.
The left and right semi-infinite leads are assumed to be in local equilibrium and their chemical potentials are equal,  $\mu_L = \mu_R = \mu$, in the absence of the bias voltage. They are shifted by a factor $qV$ when applying a bias voltage $V>0$, as  $\mu_L = \mu - qV/2$ and $\mu_R = \mu + qV/2$, where $q=-e$ ($e>0$) is the unity charge of an electron. Therefore, the electrons always flow from left to the right.
Hereinafter, natural units are taken, i.e. $e=1$, $\hbar = 1$.


\subsection{Many-body perturbation expansion}

The Hamiltonian in Eq.\eqref{eq: Hamiltonian of transport model} cannot be straightforwardly diagonalized and neither equilibrium nor non-equilibrium properties can be obtained exactly in a full extent. We hence use the many-body perturbation theory with  Green functions extended to non-equilibrium situations within the Schwinger-Keldysh formalism. Instead of the linear one-way time ordering of equilibrium we have to introduce an ordering along a Keldysh contour $\mathbb C$ in the plane of complex times consisting generally from three branches out of equilibrium \cite{Wagner:1991ug,stefanucci2013nonequilibrium}. 

Since we are interested only in the physics on the dot, the non-local degrees of freedom  of the leads can be projected (integrated) out and to a problem with only  local dynamical degrees of freedom.  We define the local Keldysh contour-ordered Green function on the quantum dot as
\begin{dmath}\label{eq: contour-Keldysh Green function}
G_\sigma(z,z') = -i \avg{ T_{\mathbb C}\left\{d_\sigma(z), d_\sigma^\dag(z')\right\} } \,,
\end{dmath}
where the thermal average of  operator $O$ is defined by $\avg{O} = Tr[\rho O]$, where $\rho$ is the equilibrium density operator at an initial time $t_i$. Further, $d_\sigma(z)$ and $d_\sigma^\dag(z')$ are the operators  in the Heisenberg picture evolving along the Keldysh contour $\mathbb C$.
The first forward branch of the Keldysh contour goes from $t_i$ to $t_f$, the second one returns back from $t_f$ to $t_i$, and the third one is purely imaginary from $t_i$ to $t_i - i\beta$. We used  a symbol  $T_{\mathbb C}$ for the contour-ordering operator which put the operators along the contour in the ascending order according to complex time $z\in \mathbb C$. 

Although we will derive the formulas in the general Keldysh formalism we will finally be interested only in long times with $t_{i}\to -\infty $ and the Hubbard interaction $U$ switched on adiabatically. The initial state at $t_{i}$ has then only little impact on the observed time evolution.  We can hence can cut off the imaginary leg of the Keldysh contour and replace it with thermal averaging with the non-interacting Hamiltonian \cite{Schwinger:1961aa,keldysh1965diagram,haug2008quantum,kamenev2011field,stefanucci2013nonequilibrium}. The Green function from Eq.~\eqref{eq: contour-Keldysh Green function} can then be represented as    
\begin{dmath}\label{eq: contour-ordered Green function}
G_\sigma(z,z') = -i\avg{T_\supset\{ d_\sigma(z), d_\sigma^\dag(z') \}} \,,
\end{dmath}
where the angular brackets denote now the thermodynamic averaging with the non-interacting Hamiltonian and the $T_{\supset}$ is the time-ordering along the two-leg Keldysh real-time contour.  One can derive various real-time Green functions  from the contour-ordered one, $G_\sigma(z,z')$, depending on which branch the time arguments $z$ and $z'$ reside, (see Appendix \ref{sec_app: real time GFs}). 

We first resolve the non-equilibrium Green function from Eq.~\eqref{eq: contour-ordered Green function} for the non-interacting dot, that is for $U=0$, exactly. 
Their impact on the physics of the dot reduces to a hybridization self-energy $\Sigma_\sigma^{ld}(z,z') = \sum_{s\in{L,R}} \Sigma_{s\sigma}^{ld}(z,z')$ after integrating their degrees of freedom.
We further use a wide-band limit (WBL) where the density of states of the lead electrons  is approximated by its value at the chemical potential \cite{PhysRevB.50.5528}.
This approximation  works well if the density of states  of the leads only slowly varies around the Fermi level, see Appendix \ref{sec_app: lead self-energy}.

The  hybridization self-energy from the leads enters the following left and right Dyson equations for the two time variables of the non-equilibrium Green function 
 \cite{haug2008quantum,stefanucci2013nonequilibrium}
\begin{subequations}\label{eq: EOM of non-interaction QD Green function}
\begin{multline}\label{eq: EOM of non-interaction QD Green function left}
\left(+i\frac{\overrightarrow{d}}{dz} - \epsilon_d + \sigma h\right) G^{0}_\sigma(z,z') 
\\
= \delta(z-z') + \int_\supset d\bar{z} \Sigma^{ld}_\sigma(z,\bar{z}) G^0_\sigma(\bar{z},z'),
\end{multline}
\begin{multline}\label{eq: EOM of non-interaction QD Green function, right}
G_\sigma^0(z,z') \left( -i\frac{\overleftarrow{d}}{dz'} - \epsilon_d + \sigma h \right) 
\\
= \delta(z-z') + \int_\supset d\bar{z} G_\sigma^0(z,\bar{z}) \Sigma_\sigma^{ld}(\bar{z},z').
\end{multline}
\end{subequations}
 Notice that $\delta(z - z')$ is a contour delta function so that $\int_{\mathbb C}dz \delta(z - z') =1$. It means that $\delta(-i(\tau - \tau')) = i\delta(\tau - \tau')$ for $\tau, \tau' \in(0,\beta)$ on the imaginary leg of the Keldysh contour. 



The impact of the Coulomb interaction is contained in the the interaction self-energy $\Sigma_\sigma^{int}(z,z')$ that will be determined from the many-body perturbation theory  \cite{bruus2004many,haug2008quantum,stefanucci2013nonequilibrium}. The interaction self-energy  $\Sigma_\sigma^{int}(z,z')$ determines the full non-equilibrium Green function from Eq.~\eqref{eq: contour-ordered Green function} via another Dyson integral equation 
\begin{multline}\label{eq: Dyson equation}
G_\sigma(z,z') = G^{0}_\sigma(z,z') 
\\
+\ \int_\supset dz_{\bar{1}} dz_{\bar{2}} G^{0}_\sigma(z,z_{\bar{1}}) \Sigma^{int}_\sigma(z_{\bar{1}},z_{\bar{2}}) G_\sigma(z_{\bar{2}},z')\,.
\end{multline}

The interacting self-energy should be calculated from the renormalized perturbation expansion in the interaction strength. A consistent scheme for introducing renormalizations in the equilibrium perturbation theory was introduced by Baym and Kadanoff \cite{Baym:1961aa,Baym:1962aa,kadanoff1962quantum}.
This scheme was later extended to non-equilibrium quantum transport \cite{PhysRevB.77.115333,PhysRevB.79.155110,haug2008quantum,stefanucci2013nonequilibrium}. 
The interacting self-energy is directly related to the two-particle vertex $\Gamma_{\sigma\bar{\sigma}}$ in the Baym-Kadanoff approach  via the Schwinger-Dyson equation. Its non-equilibrium form for the quantum dot is
\begin{multline}\label{eq: self-energy -- Keldysh contour}
	\Sigma_{\sigma}^{int}(z,z') = Un_{\bar{\sigma}}(z) \delta(z-z') \\
 - iU \int_\supset dz_{\bar{1}} dz_{\bar{2}} dz_{\bar{4}} G_\sigma(z,z_{\bar{1}}) \Gamma_{\sigma\bar{\sigma}}(z_{\bar{1}}, z_{\bar{2}}, z', z_{\bar{4}}) 
 \\ \times G_{\bar{\sigma}}(z_{\bar{4}},z) G_{\bar{\sigma}} (z, z_{\bar{2}}) \,.
\end{multline}
where $\bar{\sigma} = -\sigma$.

\begin{figure}
\includegraphics[width=1.0\linewidth]{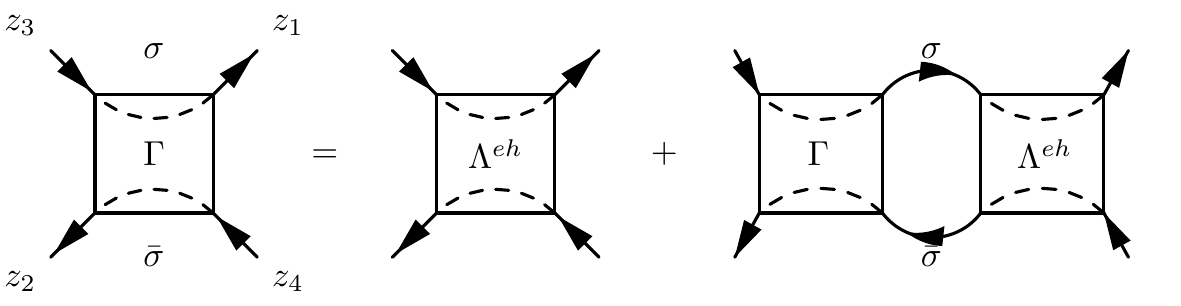}
\caption{Diagrammatic representation of the Bethe-Salpeter equation in the electron-hole channel with the notation of the vertex and the spin variables.
\label{fig:BS-eh}}
\end{figure}

The self-energy is the irreducible part of the on-electron propagator. We can also introduce a two-particle irreducible vertex from which the full vertex is obtained via a Bethe-Salpeter equation in analogy with the Dyson equation. Unlike the one-particle irreducibility the two-particle irreducibility is not uniquely defined \cite{Bickers:1991ab,Janis:1998aa,Janis:1999aa,Rohringer:2018aa}. If we choose the electron-hole irreducibility and introduce the electron-hole irreducible vertex $\Lambda^{eh}$ the non-equilibrium Bethe-Salpeter equation for the two-particle vertex reads  
\begin{widetext}
\begin{dmath}\label{eq: BSE in eh-channel -- Keldysh contour}
	\Gamma_{\sigma\bar{\sigma}}(z_1,z_2,z_3,z_4) = \Lambda_{\sigma\bar{\sigma}}^{eh}(z_1,z_2,z_3,z_4)
	+ \int_\supset dz_{\bar{1}}dz_{\bar{2}}dz_{\bar{3}}dz_{\bar{4}}\Gamma_{\sigma\bar{\sigma}}(z_{\bar{1}},z_2,z_3,z_{\bar{4}})
	G_\sigma(z_{\bar{3}},z_{\bar{1}}) G_{\bar{\sigma}}(z_{\bar{4}},z_{\bar{2}})
	 \Lambda_{\sigma\bar{\sigma}}^{eh}(z_1,z_{\bar{2}},z_{\bar{3}},z_4)\,.
\end{dmath}
\end{widetext}
Its diagrammatic representation is plotted in Fig.~\ref{fig:BS-eh}.

The two-particle irreducible vertices are not the integral part of the Baym-Kadanoff approach where only one-particle functions are used. They are, however, important for checking whether the solution with the two-particle vertex $\Gamma$  from the Schwinger-Dyson equation~\eqref{eq: self-energy -- Keldysh contour} and obeying the Bethe-Salpeter equation~\eqref{eq: BSE in eh-channel -- Keldysh contour} is conserving. It is the case if the self-energy $\Sigma$ and the two-particle irreducible vertex $\Lambda^{eh}$  obey a functional Ward identity \cite{Baym:1962aa} that in our notation and the selection of the variables of the two-particle vertex, see Fig.~\ref{fig:BS-eh},  reads   
\begin{equation}\label{eq:Ward-non-equilibrium}
 \Lambda_{\sigma\bar{\sigma}}^{eh}(z_1,z_2,z_3,z_4) = - \frac{\delta \Sigma_{\sigma}(z_{1},z_{3})}{\delta G_{\bar{\sigma}}(z_{4},z_{2})} \,.
\end{equation} 
It was shown, however, that no approximate solution can obey simultaneously the Schwinger-Dyson equation and the Ward identity with a single self-energy  and a single two-particle vertex \cite{Janis:1998aa,Janis:2019aa}. We show in the next section how to qualitatively reconcile the Ward identity and the Schwinger-Dyson equation.

\section{Two-particle self-consistency\label{sec: 2PSCF}}

\subsection{Generating two-particle vertex and self-energies}

The Baym-Kadanoff approach is based on the existence of a generating Luttinger-Ward functional $\Phi[G,U]$ from which all quantities are derived via functional derivatives with respect to $G$. The first derivative of this functional leads to the Schwinger-Dyson equation for the self-energy that is uniquely defined. Its second derivatives lead to two-particle irreducible vertices. Since we cannot obey Ward identity, Eq.~\eqref{eq:Ward-non-equilibrium} and the Schwinger-Dyson equation~\eqref{eq: self-energy -- Keldysh contour} simultaneously in any approximation with a single self-energy, two-particle vertices are then defined ambiguously. It does not matter much if the difference between the two-particle vertices from the Schwinger-Dyson equation and from the Bethe-Salpeter equation with the irreducible vertex from the Ward identity is only quantitative. We come into trouble, however, if we approach a critical point of a continuous phase transition. The phase transition remains continuous only if the Ward identity is obeyed at least in the linear order of the symmetry-breaking field, conjugate to the order parameter. The uniqueness of the critical behavior demands the existence of a unique two-particle vertex. That is why one of the authors proposed an alternative construction of the renormalized perturbation expansion. The basic idea of this construction is to use the two-particle irreducible vertex from the critical Bethe-Salpeter equation as the generating functional of the perturbation theory \cite{Janis:2007aa,Janis:2008ab,Janis:2017aa,Janis:2017ab,Janis:2019aa,Janis:2020aa}. This construction can be straightforwardly extended beyond equilibrium.     

We assume here that the potential critical behavior is a transition to a magnetically ordered phase with the order parameter conjugate to the magnetic field. The potentially divergent Bethe-Salpeter equation in equilibrium is that with multiple electron-hole scatterings, Eq.~\eqref{eq: BSE in eh-channel -- Keldysh contour}. We hence choose $\Lambda^{eh}_{\uparrow\downarrow}$ as the generating functional. It is sufficient to resolve the Ward identity only in the linear order in the magnetic field to maintain consistency between the order parameter and the singular two-particle vertex. To do so, we separate one-particle functions to those with odd and even symmetry with respect to the magnetic field. We denote 
\begin{subequations}
\begin{align}
\Delta G(z,z') &=  \ \frac 12\left[G_{\sigma}(z,z') - G_{\bar{\sigma}}(z,z') \right]\,, 
 \\
\bar{G}(z,z') &= \frac 12\left[G_{\sigma}(z,z') + G_{\bar{\sigma}}(z,z')\right] \,.
\end{align}
\end{subequations}
%
The self-energy resolved from Ward identity~\eqref{eq:Ward-non-equilibrium} for a given two-particle irreducible vertex $\Lambda^{eh}$ has odd symmetry and in the the linear order it reads 
\begin{multline}\label{eq: odd self-energy -- Keldysh contour}
	\Delta \Sigma^{int}(z_1,z_3) 
	\\
	= \int_\supset dz_2 dz_4 \Lambda^{eh}_{s}(z_1,z_2,z_3,z_4) \Delta G(z_4,z_2),
\end{multline}
where $\Lambda^{eh}_{s} = ( \Lambda^{eh}_{\downarrow\uparrow} + \Lambda^{eh}_{\uparrow\downarrow} )/2$ is the symmetric spin-singlet electron-hole irreducible vertex. We do not need to consider odd two-particle functions since the order parameter of the ordered phase is a one-particle quantity. 

The Ward identity affects only the odd self-energy being non-zero only in the ordered phase.
The even self-energy is untouched by the Ward identity and is determined separately from the Schwinger-Dyson equation. We must, however, use the symmetrized one-particle propagator $\bar{G}$ not to change the thermodynamic consistency, that is, the critical behavior of the vertex from the Schwinger-Dyson equation is that determined from the derivative of the odd self-energy. The Schwinger-Dyson equation~\eqref{eq: self-energy -- Keldysh contour} changes to
\begin{multline}\label{eq: SDE -- symmetrized}
	\bar{\Sigma}^{int}(z,z') = \frac{U}{2}n(z) \delta(z-z') 
	\\ - i U\int_\supset dz_{\bar{1}} dz_{\bar{2}} dz_{\bar{4}} \bar{G}(z,z_{\bar{1}})
	\Gamma_{s}(z_{\bar{1}}, z_{\bar{2}}, z', z_{\bar{4}}) 
	\\
	\times\bar{G}(z_{\bar{4}},z) \bar{G} (z, z_{\bar{2}})
\end{multline}
where $n(z) = n_\uparrow(z) + n_\downarrow(z)$ is the total charge density
and the Bethe-Salpeter equation~\eqref{eq: BSE in eh-channel -- Keldysh contour} changes to
\begin{multline}\label{eq: BSE in eh-channel -- symmetrized}
	\Gamma_{s}(z_1,z_2,z_3,z_4) = \Lambda^{eh}_{s}(z_1,z_2,z_3,z_4)
	\\
	+ \int_\supset dz_{\bar{1}}dz_{\bar{2}}dz_{\bar{3}}dz_{\bar{4}}\Gamma_{s}(z_{\bar{1}},z_2,z_3,z_{\bar{4}})
	\bar{G}(z_{\bar{3}},z_{\bar{1}}) 
	\\
	\bar{G}(z_{\bar{4}},z_{\bar{2}})
	 \Lambda^{eh}_{s}(z_1,z_{\bar{2}},z_{\bar{3}},z_4)\,.
\end{multline}
%
%
The full interaction spin-dependent self-energy is then a sum of the even and the odd self-energies, that is \cite{Janis:2019aa}
\begin{dmath}
\Sigma^{int}_\sigma(z,z') = \bar{\Sigma}^{int}(z,z') + \sigma \Delta\Sigma^{int}(z,z')\,.
\end{dmath}
The unique two-particle vertex determines the self-energy that obeys the Ward identity in the linear order of the symmetry-breaking field and thereby guarantees that the order parameter develops continuously from zero below the critical point of the phase transition to a magnetically ordered phase.

\subsection{Reduced parquet equations}



The two-particle approach guarantees thermodynamic qualitative consistency between the susceptibility and the order parameter. The quality of the approximation depends on the selection of the symmetric singlet electron-hole irreducible vertex $\Lambda_{s}^{eh}$. If we choose the simplest approximation $\Lambda_{s}^{eh}(z_1,z_2,z_3,z_4)=iU\delta(z_1-z_2)\delta(z_2-z_3)\delta(z_3-z_4)$, then the symmetric self-energy leads either to FLEX or RPA spin-symmetric solutions in the high-temperature phase, depending on whether the one-particle propagators are renormalized or not. These approximations fail in the strong-coupling limit and an improved two-particle vertex should be selected. We introduce a two-particle self-consistency to suppress the spurious transition to the magnetic state of the weak-coupling approximations. The most straightforward way to introduce a two-particle self-consistency is to use the parquet construction of the two-particle vertex.     

It is sufficient to use only two Bethe-Salpeter equations to provide a reliable transition from weak to strong coupling. We choose the Bethe-Salpeter equation in the electron-hole (eh) channel in this case which becomes singular at the magnetic transition. The other equation must attenuate the tendency towards the critical point. It is the Bethe-Salpeter equation in the electron-electron (ee) channel for the magnetic transition. Its symmetric version out of equilibrium reads 
\begin{multline}\label{eq: BSE in ee-channel -- Keldysh contour}
	\Gamma_{s}(z_1,z_2,z_3,z_4) = \Lambda_{s}^{ee}(z_1,z_2,z_3,z_4)
	\\
	+\ \int_\supset dz_{\bar{1}}dz_{\bar{2}}dz_{\bar{3}}dz_{\bar{4}} \Lambda_{s}^{ee}(z_1,z_2,z_{\bar{3}},z_{\bar{4}}) \bar{G}(z_{\bar{3}},z_{\bar{1}})
	\\
	\times\bar{G}(z_{\bar{4}},z_{\bar{2}})
		\Gamma_{s}(z_{\bar{1}},z_{\bar{2}},z_3,z_4)\,.
\end{multline}

The fundamental idea of the parquet approach is to use the fact that the reducible diagrams in one channel are irreducible in the other scattering channels. We denote $\mathcal{K}^{\alpha}$ the reducible vertex in channel $\alpha$, that is $\Gamma = \Lambda^{\alpha} + \mathcal{K}^{\alpha}$.  The fundamental parquet equation in the two-channel scheme can be written in either of the following two forms  
\begin{widetext}
\begin{subequations}\label{eq: Parquet equation -- Keldysh contour}
\begin{align}\label{eq: Parquet equation -- Keldysh contour-irreducible}
	\Gamma_s(z_1,z_2,z_3,z_4) &= \Lambda^{eh}_s(z_1,z_2,z_3,z_4) + \Lambda^{ee}_s(z_1,z_2,z_3,z_4) - I_s(z_1,z_2,z_3,z_4) \,,
	\\
\label{eq: Parquet equation -- Keldysh contour-reducible}
	\Gamma_s(z_1,z_2,z_3,z_4) &= \mathcal{K}^{eh}_s(z_1,z_2,z_3,z_4) + \mathcal{K}^{ee}_s(z_1,z_2,z_3,z_4) + I_s(z_1,z_2,z_3,z_4) \,.
\end{align}
\end{subequations}
\end{widetext}%
We introduced the fully two-particle irreducible vertex $I_s(z_1,z_2,z_3,z_4)$ that becomes the generator of the perturbation theory in the parquet approach. Using one of these representations of the full two-particle vertex in the Bethe-Salpeter equations~\eqref{eq: BSE in eh-channel -- symmetrized} and~\eqref{eq: BSE in ee-channel -- Keldysh contour} we obtain a set of coupled equations determining self-consistently either  irreducible  $\Lambda^{eh}_{s}$, $\Lambda^{ee}_{s}$ or reducible  $\mathcal{K}^{eh}_{s}$, $\mathcal{K}^{ee}_{s}$ vertices.

\begin{figure}
\includegraphics[width=0.8\linewidth]{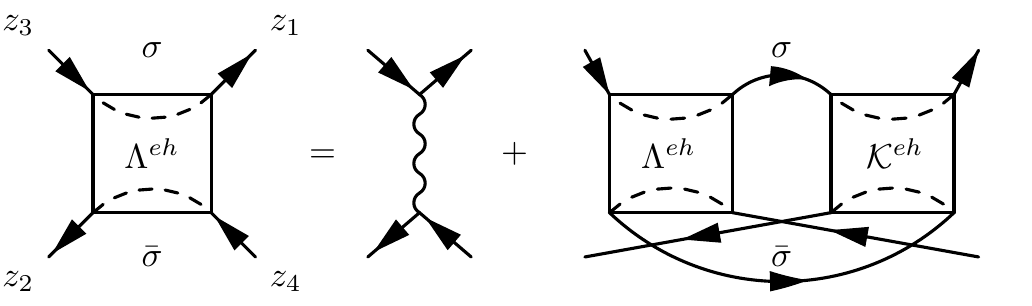}\\
\vspace{3mm}
\includegraphics[width=1.0\linewidth]{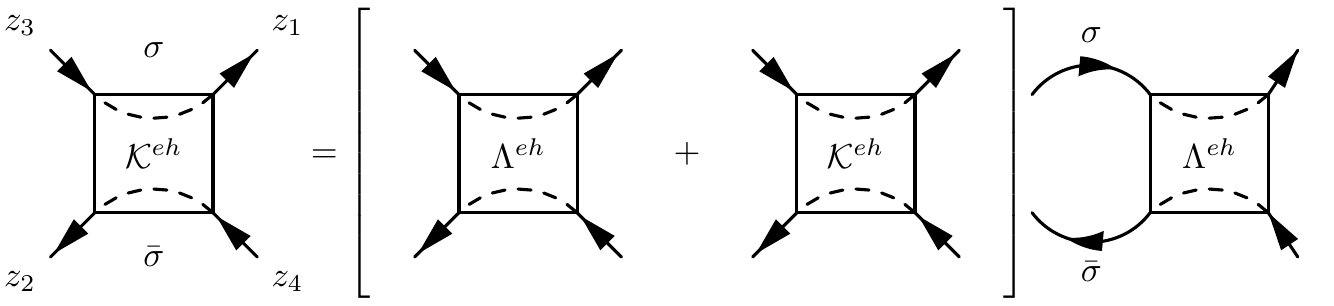}
\caption{Diagrammatic representation of the reduced Parquet equations in electron-electron (first row) and electron-hole channels (second row).
The vertex diagrams in the brackets are connected to the left propagators by the appropriate multiplication rules.
\label{fig: reduced Parquet equation}}
\end{figure}

The full solution of the parquet equations with $I_s(z_1,z_2,z_3,z_4)=iU\delta(z_1-z_2)\delta(z_2-z_3)\delta(z_3-z_4)$ and with two or three channels, suppresses the critical behavior \cite{Janis:2006ab}. One has either to go beyond the bare interaction for the fully irreducible vertex or one can modify the parquet equations so that the critical behavior of the weak-coupling approximations is not completely destroyed. One of the authors proposed to get rid of the terms in the parquet equations that suppress the critical behavior in the electron-hole channel and replaced the full set of the two-channel parquet equations with a couple of the so-called reduced parquet equations   \cite{Janis:2019aa}. The equation for the regular irreducible vertex in the electron-hole channel is then reduced to   
\begin{multline}\label{eq: RPE in ee-channel -- Keldysh contour}
\Lambda^{eh}_{s}(z_1,z_2,z_3,z_4) = iU\delta(z_1-z_3)\delta(z_2-z_4)\delta(z_1-z_4) 
\\
+\ \int_\supset dz_{\bar{1}} dz_{\bar{2}} dz_{\bar{3}} dz_{\bar{4}} \Lambda^{eh}_{s}(z_{\bar{1}},z_{\bar{2}},z_3,z_4) \bar{G}(z_{\bar{3}},z_{\bar{1}}) 
\\
\times \bar{G}(z_{\bar{4}},z_{\bar{2}}) \mathcal{K}^{eh}_{s}(z_1,z_2,z_{\bar{3}},z_{\bar{4}}).
\end{multline}
The equation for the reducible vertex in the electron-hole channel remains unchanged
\begin{multline}\label{eq: RPE in eh-channel -- Keldysh contour}
\mathcal{K}^{eh}_{s}(z_1,z_2,z_3,z_4) =  \int_\supset dz_{\bar{1}}dz_{\bar{2}}dz_{\bar{3}}dz_{\bar{4}} \\
\left[
\mathcal{K}^{eh}_{s}(z_{\bar{1}},z_2,z_3,z_{\bar{4}}) + \Lambda^{eh}_{s}(z_{\bar{1}},z_2,z_3,z_{\bar{4}}) \right]
\bar{G}(z_{\bar{3}},z_{\bar{1}}) 
\\
\times \bar{G}(z_{\bar{4}},z_{\bar{2}})
\Lambda_{s}^{eh}(z_1,z_{\bar{2}},z_{\bar{3}},z_4).
\end{multline}
The diagrammatic representation of these equations is given in Fig.\ref{fig: reduced Parquet equation}.
We showed that the reduced parquet equations correctly reproduce the Kondo regime of SIAM at equilibrium.\cite{Janis:2017aa,Janis:2019aa,Janis:2020ab}

\subsection{Instantaneous effective interaction}

The reduced parquet equations do not simplify the complexity of the full set of parquet equations. They cannot be solved easily due to the unrestricted four-time dependence of the two-particle vertices.
To simplify the problem, we resort to an instantaneous effective interaction or the  local time approximation (LTA), which assumes that the  irreducible vertex $\Lambda^{eh}_{s}(z_1,z_2,z_3,z_4)$ reduces to an instantaneous effective interaction  $i\tilde{\Lambda}_{s}^{eh}(z_4)\delta(z_1-z_3)\delta(z_2-z_4)\delta(z_1-z_4)$.  The reducible vertex $\mathcal{K}^{eh}_{s}(z_1,z_2,z_3,z_4)$ thus is partially diagonalized, having non-zero values only when $z_1=z_4$ and $z_2=z_3$, thus one can define $\mathcal{K}^{eh}_{s}(z_1,z_2,z_3,z_4) = i\tilde{\mathcal{K}}^{eh}_{s}(z_1,z_2)\delta(z_1-z_4)\delta(z_2-z_3)$.
As a result, the reduced parquet equations, Eq.\eqref{eq: RPE in ee-channel -- Keldysh contour} and Eq.\eqref{eq: RPE in eh-channel -- Keldysh contour}, simplify to
\begin{multline}\label{eq: RPE in ee-channel with static decoupling -- Keldysh contour}
\left[\tilde{\Lambda}_{s}^{eh}(z_4) - U\right] \delta(z_1-z_4)\delta(z_2-z_4)
\\
= i\tilde{\Lambda}_{s}^{eh}(z_4) \tilde{\mathcal{K}}^{eh}_{s}(z_1,z_2)\bar{G}(z_2,z_4)\bar{G}(z_1,z_4) \,,
\end{multline}
\begin{multline}\label{eq: RPE in eh-channel with static decoupling -- Keldysh contour}
-\left[\tilde{\Lambda}_{s}^{eh}(z_4)\right]^2 \phi_{s}(z_1,z_2)
=\tilde{\mathcal{K}}_{s}^{eh}(z_1,z_2) 
\\
+\ \tilde{\Lambda}_{s}^{eh}(z_4) \int_\supset dz_{\bar{4}} \phi_{s}(z_1,z_{\bar{4}}) \tilde{\mathcal{K}}_{s}^{eh}(z_{\bar{4}},z_2)\,,
\end{multline}
where we introduced the singlet electron-hole bubble
\begin{dmath}\label{eq: electron-hole bubble -- Keldysh contour}
\phi_{s}(z_1,z_2) = -i \bar{G}(z_1,z_2) \bar{G}(z_2,z_1)\,.
\end{dmath}
%

Equation~\eqref{eq: RPE in eh-channel with static decoupling -- Keldysh contour} can be resolved for the reducible vertex $\tilde{\mathcal{K}}_{s}^{eh}(z_1,z_2) $. If we insert its solution into Eq.~\eqref{eq: RPE in ee-channel with static decoupling -- Keldysh contour} its right-hand side maintains a non-trivial dependence on complex times $z_{1}$ and $z_{2}$. Since the left-hand side does not depend on $z_{1},z_{2}$, we have to resign on point-wise equality in Eq.~\eqref{eq: RPE in ee-channel with static decoupling -- Keldysh contour} when the irreducible vertex is approximated via an instantaneous effective interaction. Instead, we  replace then Eq.\eqref{eq: RPE in ee-channel with static decoupling -- Keldysh contour} with and equality where both sides are averaged over the redundant time variables. Various averaging schemes can be applied, see \cite{Janis:2019aa,PhysRevB.103.235163,doi:10.1063/9.0000019}.
Different schemes only quantitatively affect the physical behavior far away from the critical region but do  qualitatively change the critical behavior itself. Here we  multiply both sides of  Eq.~\eqref{eq: RPE in ee-channel with static decoupling -- Keldysh contour} by $\bar{G}(z_4,z^+_2) \bar{G}(z_4,z^+_1)$  and integrate over the redundant time  variables $z_1$ and $z_2$. Notice that $z^+=z+0^+$ indicates an infinitesimally positive time shift of the variable $z$.
As a result, $\tilde{\Lambda}_{s}^{eh}$ can be consistently determined from the following alternative (mean-field) equation
\begin{dmath}\label{eq: RPE for irreducible vertex -- Keldysh contour}
\tilde{\Lambda}_{s}^{eh}(z) = \frac{Un(z)n(z)}{n(z)n(z) - 4 R_{s}^{eh}(z)},
\end{dmath}
with the screening integral 
\begin{multline}\label{eq: RPE for renormalization integral -- Keldysh contour}
R^{eh}_{s}(z)
\\
 = i\int_\supset d z_1 d z_2 \phi_{s}(z,z_1) \tilde{\mathcal{K}}^{eh}_{s}(z_1,z_2) \phi_{s}(z_2,z).
\end{multline}
Equations~\eqref{eq: RPE in eh-channel with static decoupling -- Keldysh contour},~\eqref{eq: RPE for irreducible vertex -- Keldysh contour} and~\eqref{eq: RPE for renormalization integral -- Keldysh contour} form a closed set  determining the two-particle vertices within local-time approximation.

The symmetric two-particle vertex simplifies to $\Gamma_{s}(z_1,z_2,z_3,z_4) = i\tilde{\Gamma}_{s}(z_1,z_2)\delta(z_1-z_4)\delta(z_2-z_3)$ where $\tilde{\Gamma}_{s}(z_1,z_2) = \tilde{\mathcal{K}}_{s}(z_1,z_2) + \tilde{\Lambda}^{eh}_{s}(z_1)\delta(z_1-z_2)$.
As result, Schwinger-Dyson equation~\eqref{eq: SDE -- symmetrized} turns to
\begin{dmath}\label{eq: even self-energy with LTA -- Keldysh contour}
\Sigma^{int}(z,z') = \frac{U}{2}n(z)\delta(z-z')
+iU \bar{G}(z,z') \int_{\supset} d\bar{z} \phi_{s}(z,\bar{z}) \tilde{\Gamma}_{s}(\bar{z},z'),
\end{dmath}
and the Ward identity, Eq\eqref{eq: odd self-energy -- Keldysh contour}, becomes
\begin{dmath}\label{eq: odd self-energy with LTA -- Keldysh contour}
\Delta \Sigma(z,z') =  i\tilde{\Lambda}^{eh}_s(z)\Delta G(z,z^+)\delta(z-z').
\end{dmath}
%
Notice that the electron-hole bubble satisfies a symmetry relation $\phi_{s}(z_1,z_2) = \phi_{s}(z_2,z_1)$ and, consequently,   $\tilde{\mathcal{K}}^{eh}_{s}(z_1,z_2) = \tilde{\mathcal{K}}^{eh}_{s}(z_2,z_1)$ and $\tilde{\Gamma}_{s}(z_1,z_2) = \tilde{\Gamma}_{s}(z_2,z_1)$.  We drop superscript `eh'  and subscript `s'  in the following sections, since both irreducible and reducible vertices are from the same channel.

\section{Steady-state Quantum Transport\label{sec: transport}}

We now apply our two-particle construction with the reduced parquet equations and the instantaneous interaction to the steady-state quantum transport where the system is assumed to be evolved for a sufficiently long time in which it reaches a steady-state with time-independent densities and currents \cite{haug2008quantum}.

\subsection{Real-time steady-state formalism}

The steady-state equations formulated on the Keldysh contour with complex times can be transformed  to real times via  Langreth rules \cite{haug2008quantum}. We can then use a Fourier transform from time to real frequencies making the defining equations out of equilibrium close to those  from equilibrium.
Eq.~\eqref{eq: RPE for irreducible vertex -- Keldysh contour} becomes time-independent 
\begin{dmath}\label{eq: RPE for irreducible vertex -- steady-state}
\tilde{\Lambda} = \frac{U n^{2}}{n^{2} - 4R}\,,
\end{dmath}
since the irreducible vertex and the electron density become time-independent in the steady-state case.
The screening integral in Eq.\eqref{eq: RPE for irreducible vertex -- steady-state} is (we refer to Appendix \ref{sec_app: renormalization integral} for details)

\begin{multline}\label{eq: renormalization integral -- steady-state}
R = \frac{i}{4\pi} \int_{-\infty}^\infty dx \left\{ \left[ \phi^<(x) + \phi^>(x) \right] \tilde{\mathcal{K}}^{a}(x)\phi^a(x) 
\right. \\ \left.
+\ \phi^r(x) \left[ \tilde{\mathcal{K}}^{<}(x) + \tilde{\mathcal{K}}^{>}(x) \right]\phi^a(x) 
\right. \\ \left.
+\ \phi^r(x)\tilde{\mathcal{K}}^{r}(x)\left[ \phi^<(x) + \phi^>(x) \right] \right\},
\end{multline}
where superscripts $r$, $a$, $<$ and $>$ denote the retarded, advanced, lesser and greater counterparts of the real-time quantities, respectively. (see Appendix \ref{sec_app: real time GFs})
The real-time components of the electron-hole bubble satisfy the relations $\phi^{r/a}(x) = \phi^{a/r}(-x)$ and $\phi^\lessgtr(x) = \phi^\gtrless(-x)$ and the same holds for the real-time components of $\tilde{\mathcal{K}}$.
Explicitly, we have from Eq.\eqref{eq: electron-hole bubble -- Keldysh contour} 
\begin{subequations}
\begin{multline}
	\phi^{r/a}(w)
	= \frac{-i}{2\pi}\int_{-\infty}^\infty dx
	\left[ \bar{G}(x) \bar{G}^{a/r}(x-w) 
\right. \\ \left.	
	+\ \bar{G}^{r/a}(x) \bar{G}^<(x-w) \right]\,,
\end{multline}
\begin{dmath}
	\phi^\lessgtr(w) = \frac{-i}{2\pi}\int_{-\infty}^\infty dx \bar{G}^\lessgtr(x) \bar{G}^\gtrless
(x-w)\,.
\end{dmath}
\end{subequations}
We further obtain from Eq.\eqref{eq: RPE in eh-channel with static decoupling -- Keldysh contour}
\begin{subequations}
\begin{dmath}
\tilde{\mathcal{K}}^{r/a}(w) + \tilde{\Lambda} = \frac{\tilde{\Lambda}}{1+\tilde{\Lambda}\phi^{r/a}(w)}\,,
\end{dmath}
\begin{dmath}
\tilde{\mathcal{K}}^\lessgtr(w) 
= -  [\tilde{\Lambda} + \tilde{\mathcal{K}}^{r}(w)] \phi^\lessgtr(w) [\tilde{\Lambda} + \tilde{\mathcal{K}}^{a}(w)]\,.
\end{dmath}
\end{subequations}
With the above equations, one can self-consistently calculate the two-particle vertices with the given one-particle Green functions.

Once the two-particle vertices are determined, the even and odd parts of the self-energies can be calculated from the Schwinger-Dyson equation and Ward identity, respectively.
In particular, the real-time even self-energies read $\bar{\Sigma}^{int,r/a}(w) = Un/2 + \bar{\Sigma}^{cor,r/a}(w)$ and $\bar{\Sigma}^{int,\lessgtr}(w) = \bar{\Sigma}^{cor,\lessgtr}(w)$, where, from Eq.\eqref{eq: even self-energy with LTA -- Keldysh contour}, we have
\begin{widetext}
\begin{subequations}
\begin{multline}\label{eq: even self-energy r/a -- real frequency}
\bar{\Sigma}^{cor,r/a}(w) = \frac{iU\tilde{\Lambda}}{4\pi}
\\
\times\int_{-\infty}^\infty dx \frac{\bar{G}^{r/a}(x)\left[\phi^<(w-x) + \phi^>(w-x) \right] + \left[\bar{G}^<(x) + \bar{G}^>(x) \right]\phi^{r/a}(w-x)\left[ 1 + \tilde{\Lambda}\phi^{a/r}(w-x) \right]}{[1+\tilde{\Lambda}\phi^r(w-x)][1+\tilde{\Lambda}\phi^a(w-x)]}\,,
\end{multline}
\begin{dmath}\label{eq: eq: even self-energy lessgtr -- real frequency}
\bar{\Sigma}^{cor,\lessgtr}(w) = \frac{iU\tilde{\Lambda}}{2\pi} \int_{-\infty}^\infty dx \frac{\bar{G}^\lessgtr(x) \phi^\lessgtr(w-x)}{[1+\tilde{\Lambda}\phi^r(w-x)][1+\tilde{\Lambda}\phi^a(w-x)]}\,.
\end{dmath}
\end{subequations}
\end{widetext}
The real time odd self-energies, from Eq.\eqref{eq: odd self-energy with LTA -- Keldysh contour}, become
\begin{dmath}\label{eq: odd self-energy r/a -- real frequency}
	\Delta\Sigma^{int,r/a}(w) = \frac{i\tilde{\Lambda}}{2\pi} \int_{-\infty}^\infty dx \Delta G^<(x) = -\frac{m}{2}\tilde{\Lambda},
\end{dmath}
and $\Delta \Sigma^{int,\lessgtr}(w) = 0$.
The total self-energies are given by $\Sigma_\sigma^{int,x}(w) = \bar{\Sigma}^{int,x}(w) + \sigma \Delta\Sigma^{int,x}(w)$ where $x = r, a, >, <$.
The spin-dependent renormalized one-particle Green functions are
\begin{subequations}
\begin{multline}
G^{r/a}_\sigma(w)
\\
= \frac{1}{w + \sigma h_\Lambda - \epsilon_U - \Sigma^{ld,r/a}_\sigma(w) - \bar{\Sigma}^{cor,r/a}(w)},
\end{multline}
\begin{equation}
G_\sigma^\lessgtr(w) = G_\sigma^r(w)\left[ \Sigma_\sigma^{ld,\lessgtr}(w) + \bar{\Sigma}^{cor,\lessgtr}(w) \right] G_\sigma^a(w),
\end{equation}
\end{subequations}
where $\epsilon_U = \epsilon_d + Un/2$, $h_\Lambda = h + \tilde{\Lambda}m/2$.
The second equation is also well-known as the Keldysh formula.\cite{haug2008quantum}\footnote{We neglected the bound state contributions which is irrelevant to the steady-state transport.}
%

The renormalized spin-dependent propagator determines the physical quantities. For example, the spin-resolved electron density reads
\begin{dmath}\label{eq: spin-resolved density}
n_\sigma = -\frac{i}{2\pi} \int_{-\infty}^\infty G_\sigma^<(x) dx.
\end{dmath}
The total electron density and the magnetization are then given by $n = n_\uparrow + n_\downarrow$ and $m = n_\uparrow - n_\downarrow$.
The spin-resolved current go through $s$-lead is given by\cite{haug2008quantum}
\begin{multline}\label{eq: frequency-resolved current formula}
	J_{s\sigma} = \frac{q}{2\pi} 
	\\
	\times\int_{-\infty}^{\infty} dw \left[ \Sigma_{s\sigma}^{ld,>}(w) G_\sigma^<(w) - \Sigma_{s\sigma}^{ld,<}(w) G_\sigma^>(w) \right].
\end{multline}

\subsection{Thermodynamic and spectral calculations}


The two-particle scheme contains two sets of self-consistent equations. One set is used to determine the two-particle vertex, or the effective interaction. The other set  is used to determine the self-energies from the two-particle vertex. One-particle propagators are used in both sets.  They are an input in the parquet equations. The vertex function  from the parquet equations determines the thermodynamic response and controls the critical behavior of the equilibrium solution. The only consistency condition between the one-particle propagators and the two-particle vertex there is that the odd self-energy of the propagators is determined from the irreducible vertex via the Ward identity. There is no restriction on the even self-energy in the parquet equations.   We can hence separate the one and two-particle self-consistencies. We introduce thermodynamic Green functions when we use only the static (HF mean-field) even self-energy in the one-particle propagators determining the two-particle vertex. We will call the Green functions with the full even self-energy from the Schwinger-Dyson equation spectral propagators.

Hereinafter, we denote the quantities calculated with thermodynamic (spectral) calculation by superscript `$T$ ($S$)', respectively. The thermodynamic Green functions are
\begin{subequations}
\begin{dmath}
G^{T,r/a}_\sigma(w) = \frac{1}{w+\sigma h^T_\Lambda-\epsilon^T_U-\Sigma_\sigma^{ld,r/a}(w)}\,,
\end{dmath}
\begin{dmath}
G^{T,\lessgtr}_\sigma(w) = G_\sigma^{T,r}(w)\Sigma_\sigma^{ld,\lessgtr}(w) G_\sigma^{T,a}(w)\,,
\end{dmath}
\end{subequations}
where $h^T_U = h + \tilde{\Lambda}m^T /2$ and $\epsilon^T_U = \epsilon +  Un^T/2$.
Here, $n^T$ and $m^T$ are calculated from the spin-resolved thermodynamic Green functions.
Furthermore, the thermodynamic susceptibility can be derived by taking the derivative w.r.t. the magnetic field
\begin{dmath}\label{eq: magnetic susceptibility in thermodynamic calculation}
	\chi^T = \frac{-\sum_\sigma \phi^{T,r}_{\sigma\sigma}(0)}{1 + \frac{1}{2}\tilde{\Lambda}\sum_\sigma\phi_{\sigma\sigma}^{T,r}(0)}\,,
\end{dmath}
where
\begin{multline}
\phi^{T,r}_{\sigma\sigma}(0) = \frac{-i}{2\pi}\int_{-\infty}^\infty dx \left[ G_\sigma^{T,<}(x) G_\sigma^{T,a}(x) 
\right.\\ \left.
+\ G_\sigma^{T,r}(x) G_\sigma^{T,<}(x) \right]\,.
\end{multline}
The one-particle quantities reduce to the Hartree ones.

The dynamical corrections are added  after determining  the two-particle vertex via the spectral symmetric self-energy from the Schwinger-Dyson equation. The spectral Green functions reads
\begin{subequations}
\begin{multline}
	G^{S,r/a}_\sigma(w) 
	\\
	= \frac{1}{w + \sigma h^T_\Lambda - \epsilon^S_U-\Sigma_\sigma^{ld,r/a}(w) - \bar{\Sigma}^{cor,r/a}(w)}\,,
\end{multline}
\begin{multline}
	G^{S,\lessgtr}_\sigma(w) 
	\\
	= G_\sigma^{S,r}(w) \left[ \Sigma_\sigma^{ld,\lessgtr}(w) + \bar{\Sigma}^{cor,\lessgtr}(w)\right] G_\sigma^{S,a}(w) \,,
\end{multline}
\end{subequations}
where $\epsilon_U^S = \epsilon +  Un^S/2$.
Here, $\epsilon_U^S$ is iterated during the spectral calculation while $h_\Lambda^T =h + \tilde{\Lambda} m^{T}/2$ is taken from the thermodynamic calculation of the two-particle vertex. In such a way the magnetic response remains qualitatively unchanged in the spectral calculations.
We stress again that the physical and measurable quantities are determined from the spectral calculations that include the dynamic correlations.

\section{Results and Discussions\label{sec: numerics}}

The reduced parquet equations be solved in full generality only numerically.  They can be solved analytically in the Kondo strong-coupling limit at half filling where the equilibrium solution approaches a critical point. We start with this limit before we analyze the general situation of the steady-state current.

\subsection{Logarithmic scaling in the strong-coupling limit}

The SIAM at zero temperature and in the electron-hole symmetric case approaches a critical point in the absence of both magnetic field and bias with increasing interaction strength. We can safely suppress all non-critical fluctuations, which allows us to find an analytic representations of the vanishing Kondo scale.   
The screening integral, Eq.\eqref{eq: renormalization integral -- steady-state},  can be simplified in this regime to (see Appendix \ref{sec_app: equilibrium formulae})
\begin{dmath}
R = -\tilde{\Lambda}^2 \int_{-\infty}^0 \Im \frac{[\phi^r(x)]^3}{1+\tilde{\Lambda}\phi^r(x)} dx,
\end{dmath}
where we used the fluctuation dissipation theorem.

We denote the denominator of the integrand in the screening integral  $D^r(x) = 1 + \tilde{\Lambda} \phi^r(x)$. Its static value $D^r(0)$  determines the dimensionless Kondo scale $a_K = D^r(0) = 1 + \tilde{\Lambda}\phi^r(0)$  and vanishes at the critical point.
It is proportional to the inverse susceptibility, see Eq.~\eqref{eq: magnetic susceptibility in thermodynamic calculation}.   We define the Kondo scale from a two-particle thermodynamic function, the inverse magnetic susceptibility. An alternative one-particle spectral definition uses the half-width at half-maximum (HWHM) of the Kondo-Abrikosov-Suhl quasiparticle peak. The advantage of the thermodynamic definition is that it can be determined analytically.

We expand the denominator function $D^r(x)\approx a_K + \dot{D}^r(0) x$ in small frequencies and keep only the first term that is dominant in the critical region with $a_{K}\to 0$. The over-dot refers to first-order derivative w.r.t. frequency.
The derivative $\dot{D}^r(0)$ is generally a complex number that at half-filling and without magnetic field becomes purely imaginary. One can explicitly evaluate the real and imaginary parts of $\dot{D}^r(0)$ at zero temperature
\begin{subequations}
\begin{multline}
\Re \dot{D}^r(0) = \frac{\tilde{\Lambda}}{\pi}\int_{-\infty}^0 dx \left[ \Re \dot{\bar{G}}^r(x) \Im \bar{G}^r(x) 
\right. \\ \left.
-\ \Re \dot{\bar{G}}^r(x) \Im \bar{G}^r(x) \right]\,,
\end{multline}
\begin{equation}
\Im \dot{D}^r(0) = - \frac{\tilde{\Lambda}}{\pi} \Im \bar{G}^r(0) \Im \bar{G}^r(0)\,.
\end{equation}
\end{subequations}
We obtain by putting the above representations into the screening integral and taking into account Eq.\eqref{eq: RPE for irreducible vertex -- steady-state}
\begin{dmath}
\tilde{\Lambda} = \frac{U n^{2}}{n^{2} - 4\tilde{\Lambda}^2 [\phi^r(0)]^3 \frac{\Im \dot{D}^r(0)}{|\dot{D}^r(0)|^2} \log a_K}\,.
\end{dmath}
Realizing  that $\tilde{\Lambda} = -1/\phi^r(0)$, $n = 1$, and $\Re \dot{D}^r(0) = 0$ in the Kondo strong-coupling regime regime we obtain an explicit analytic solution for the Kondo scale
\begin{equation}
\log a_K = \frac{U}{4} \Im \dot{D}^r(0)\,.
\end{equation}
We used the thermodynamic propagators to calculate the electron-hole bubble and hence  $\log a_K = - U/(4\Gamma)$.
We therby reproduced the linear logarithmic scaling of $a_K$ with increasing of $U$ \cite{hewson1997kondo}.
The non-universal prefactor $1/4$ in our theory slightly differs  from  $\pi/8$ of the exact solution from the Bethe ansatz \cite{hewson1997kondo}. We stress, however, that the scaling coefficient depends on the averaging scheme used in the reduced parquet equations as well as on the density of states on the dot.

\subsection{I-V characteristics in Coulomb blockade regime}

We now turn to the Coulomb blockade regime to calculate the current-voltage (I-V) characteristics of the steady-state.
We choose $U=40\Gamma$ and $T=0.1\Gamma$.
The corresponding Kondo temperature can be estimated by Haldane's formula \cite{PhysRevLett.40.416}
\begin{dmath}\label{eq: Kondo temperature}
	T_K \approx \frac{\sqrt{2\Gamma U}}{2} \exp{\frac{\pi\epsilon_d(\epsilon_d+U)}{2\Gamma U}}.
\end{dmath}
The value of the Kondo temperature with our parameters is $T_K \approx 10^{-6}\Gamma$ that is much smaller than the real temperature $T=0.1\Gamma$ and can be safely neglected. We compare the solution obtained from the reduced parquet equation with the Hartree-Fock (HF) mean-field solution that is free of dynamical correlations and also a widely used GW approximation, where the dynamical correlations are added via the nonlocal screening effect due to the electron-hole pairing.\cite{PhysRevB.77.115333,PhysRevB.79.155110,PhysRevB.77.045119}.
\begin{figure}
	\includegraphics[width=1.0\linewidth]{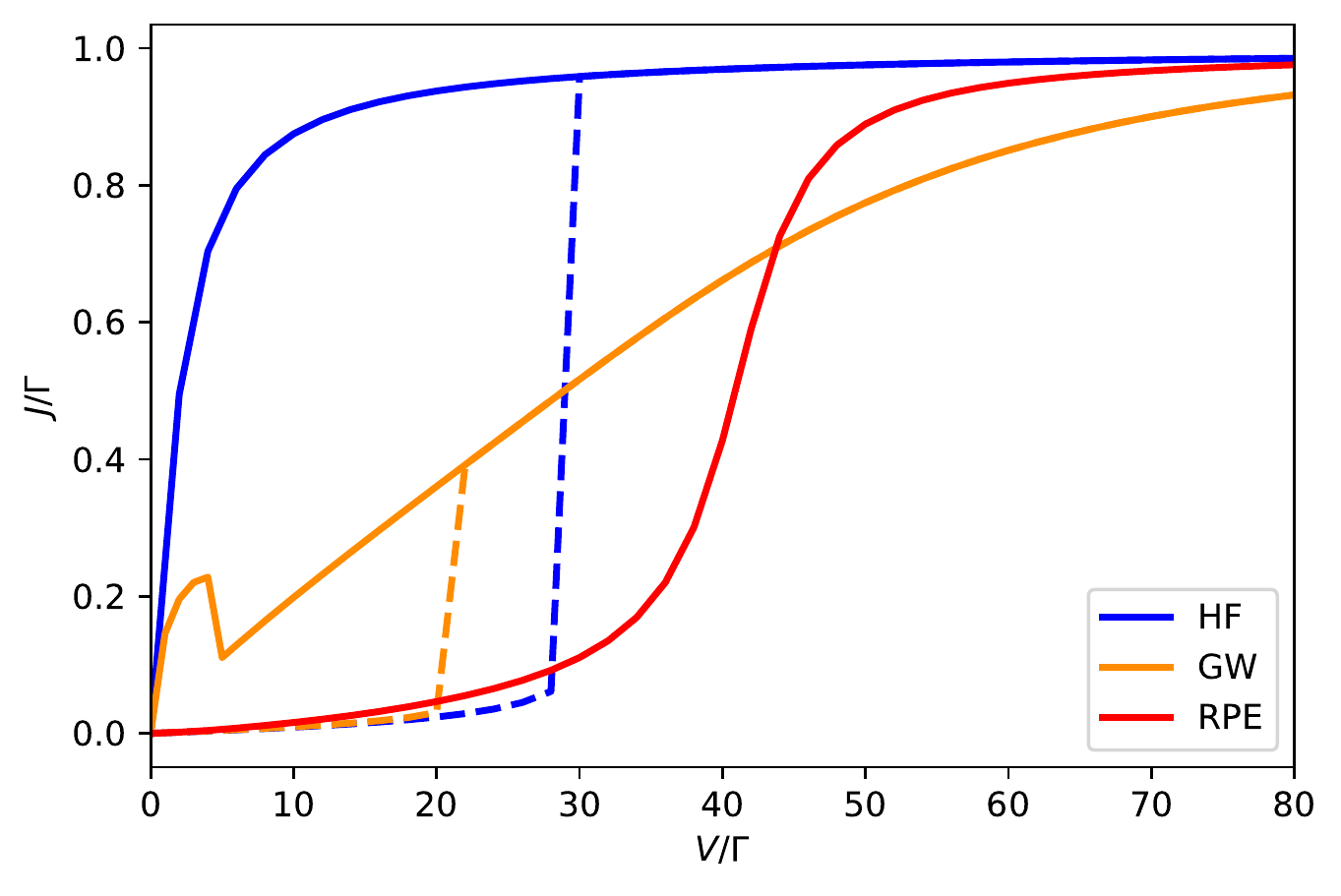}
	\caption{I-V characteristic curve of the SIAM at half-filling for $U=40\Gamma$ and $T=0.1\Gamma$ in the Coulomb-blockade regime calculated by the Hartree-Fock (HF) mean-field, GW approximation and the reduced parquet equations (RPE). Unphysical hysteresis loop appears in the HF and GW approximations with the solid line corresponding to the non-magnetic solution and the dashed line to the magnetic one. The RPE suppresses the spurious magnetic order, hence is free of the hysteresis. \label{fig: CB_IVCurve}}
\end{figure}
We plotted the I-V characteristic curve calculated by the Hatrtee-Fock (blue line), the self-consistent GW approximation (orange line) and the reduced parquet equations (red line), respectively, in Fig.~\ref{fig: CB_IVCurve}. The dynamical correlations are completely neglected in the HF mean-field and both magnetic (dash line) and non-magnetic (solid line) solutions coexist for $V < 30\Gamma$. The current in the non-magnetic state starts to grow rapidly up to saturation at a larger value of the voltage bias. The current in the magnetic state is strongly suppressed for small biases up to a (spurious) first-order transition to a non-magnetic state at a threshold value around $V=30\Gamma$. The magnetic solution in the GW approximation behaves similarly to the Hartree.-Fock one where the unphysical magnetic solution is not suppressed and the magnetic solution exists for $V < 22\Gamma$. The rapid growth of the non-magnetic curve is, however, interrupted with a small hump followed by a less steep increase towards saturation. The current in the RPE follows for weak biases the magnetic solution of the other approximations. The reduced parquet equations lead only to a non-magnetic state. The spurious magnetic transition with a discrete jump is suppreassed. Instead, it continuously crosses over to saturation as determined numerically and experimentally \cite{doi:10.1021/acs.jpcc.9b04132,Zhang:2021vy}.
\begin{figure}
	\includegraphics[width=1.0\linewidth]{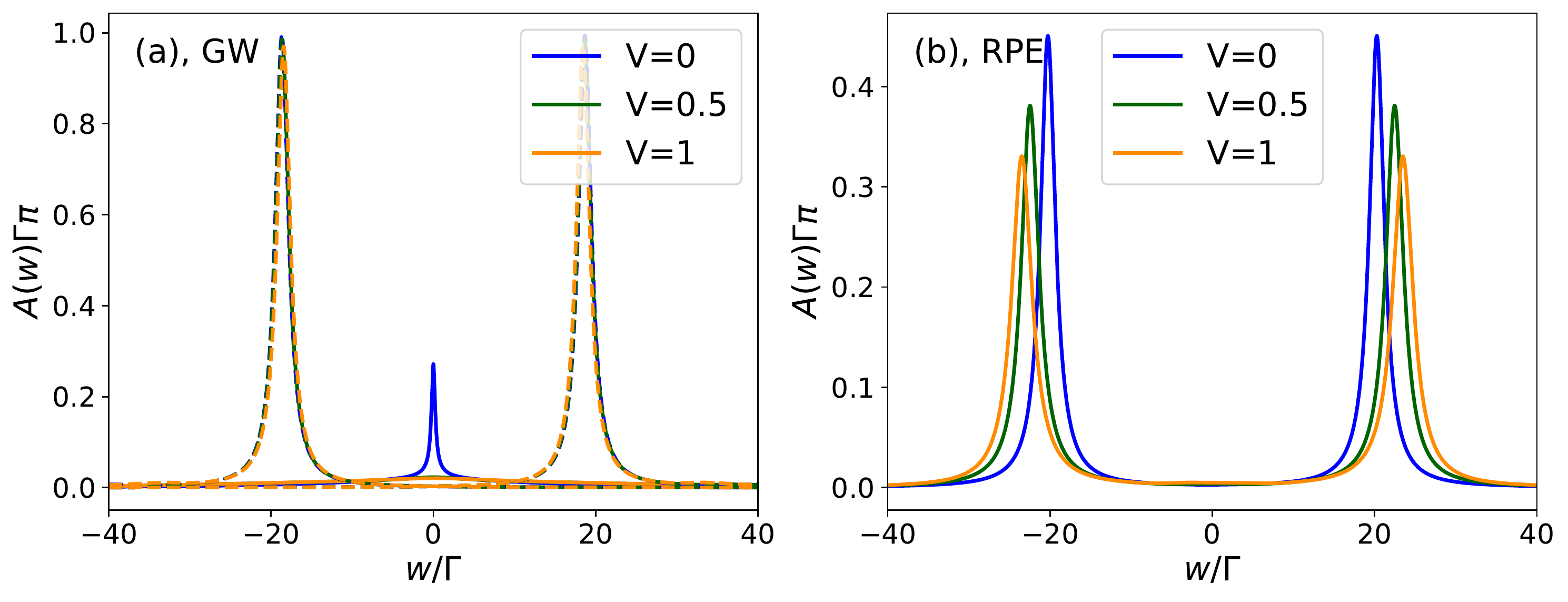}
	\caption{Spectral function for various voltage biases in GW (left panel) and the reduced parquet equations (right panel) of the SIAM at half filling, $U=40\Gamma$, and $T=0.1\Gamma\gg T_{K}$. The dashed line corresponds to the magnetic solution while the solid line for the non-magnetic one.
\label{fig: CB_spc}}
\end{figure}

The behavior of the current in the approximate solutions can be explained from the corresponding spectral functions displayed in Fig.~\ref{fig: CB_spc}. The steep increase of the current with the increasing bias of non-magnetic HF and GW solutions is caused by a central peak at the Fermi level, which is insensible to a weak applied voltage. The magnetic solution, where the spin-up and spin-down spectra are split, has vary low density of states near the Fermi level. There are hence only few electrons to participate in the current. The magnetic solution ceases to exist at larger values of the bias and the transition to the non-magnetic state leads to a jump in the current. 

The RPE solution is nonmagnetic for any bias. There is no central peak in the spectral function for temperatures high above the Kondo one. It contains two Hubbard satellite bands, sitting around $\pm U/2$ around the Fermi level. With the increasing bias, these two satellite bands do not move too much, and this leads to the `S'-shape I-V characteristics \cite{doi:10.1021/acs.jpcc.9b04132}. In particular, when $V \ll U$, the current is largely suppressed due to the small density of states around the Fermi level, exhibiting a strong Coulomb-blockade effect. The satellite bands start to contribute when the voltage becomes of the order of the position of the satellite Hubbard bands. The RPE result fits well with the magnetic solution of HF and GW at small $V$ region, due to the similar structure of the spectral functions, however, the magnetic solution breaks the spin symmetry which is unphysical.
If we further increase $V$, the I-V curve will become flat again due to the decay of the density of states at large frequencies.

\begin{figure}
	\includegraphics[width=1.0\linewidth]{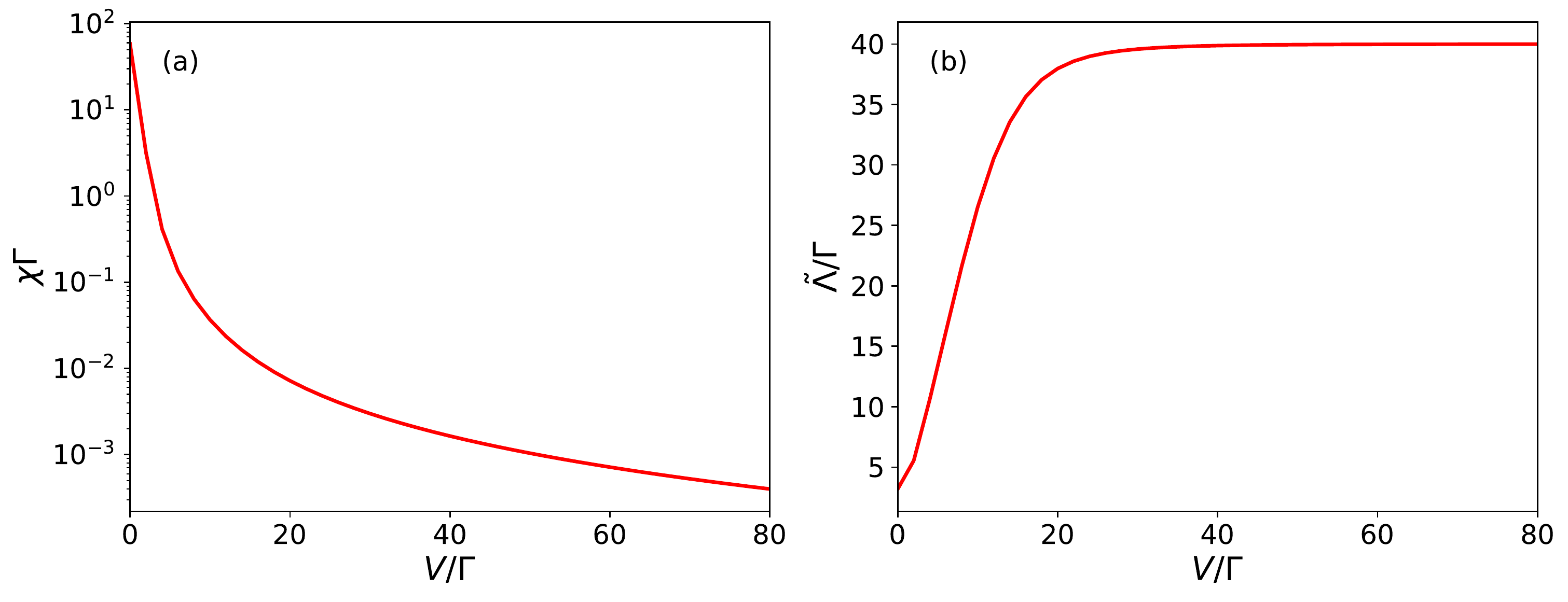}
	\caption{Magnetic susceptibility (left panel) and the effective interaction $\tilde{\Lambda}$ (right panel) of the SIAM at half filling, $U=40\Gamma$, and $T=0.1\Gamma$ as a function of the voltage bias. The susceptibility is positive and decreases with the bias while the effective interaction increases towards the bare interaction indicating that the bias suppresses the interaction-induced dynamical fluctuations. \label{fig:RPE-susceptibility}}
\end{figure}

The stability of the non-magnetic solution of the reduced parquet equations is demonstrated on the positivity of the magnetic susceptibility, Eq.~\eqref{eq: magnetic susceptibility in thermodynamic calculation}, plotted in Fig.~\ref{fig:RPE-susceptibility}. It is achieved by the two-particle self-consistency in the equation for the effective interaction, Eq.~\eqref{eq: RPE for irreducible vertex -- steady-state}. The interaction is strongly screened for small biases, the right panel of Fig.~\ref{fig:RPE-susceptibility}. The voltage suppresses the interaction-induced dynamical fluctuations and the screening of the interaction. Consequently, vertex $\tilde{\Lambda}$ approaches the bare interaction and the susceptibility exponentially decreases with increasing the bias voltage $V$. The susceptibility remains positive in the whole range of the bias voltage as expected \cite{Dirks_2013}. 

To summarize, all three approximations we compared predict the Coulomb blockade effects. The HF and GW approximations do that at the expense of the spurious magnetic transition at a small voltage.
Only approximations with a two-particle self-consistency can suppress the spurious magnetic transition and those with satellite Hubbard bands correctly reproduce the `S'-shape Coulomb-blockade I-V curve.

\subsection{Temperature and bias dependent conductance}

We now turn to study the interplay between the Kondo and Coulomb blockade effects with reduced parquet equations.
Both temperature and the biased voltage should be kept sufficiently low compared with the Kondo temperature $T_K$ to stay inside the Kondo regime.
Specifically, for the case we study below, we choose $U=6\Gamma$ with $T_K \approx 0.164\Gamma$ as estimated  from Eq.~\eqref{eq: Kondo temperature}.
\begin{figure}
\includegraphics[width=1.0\linewidth]{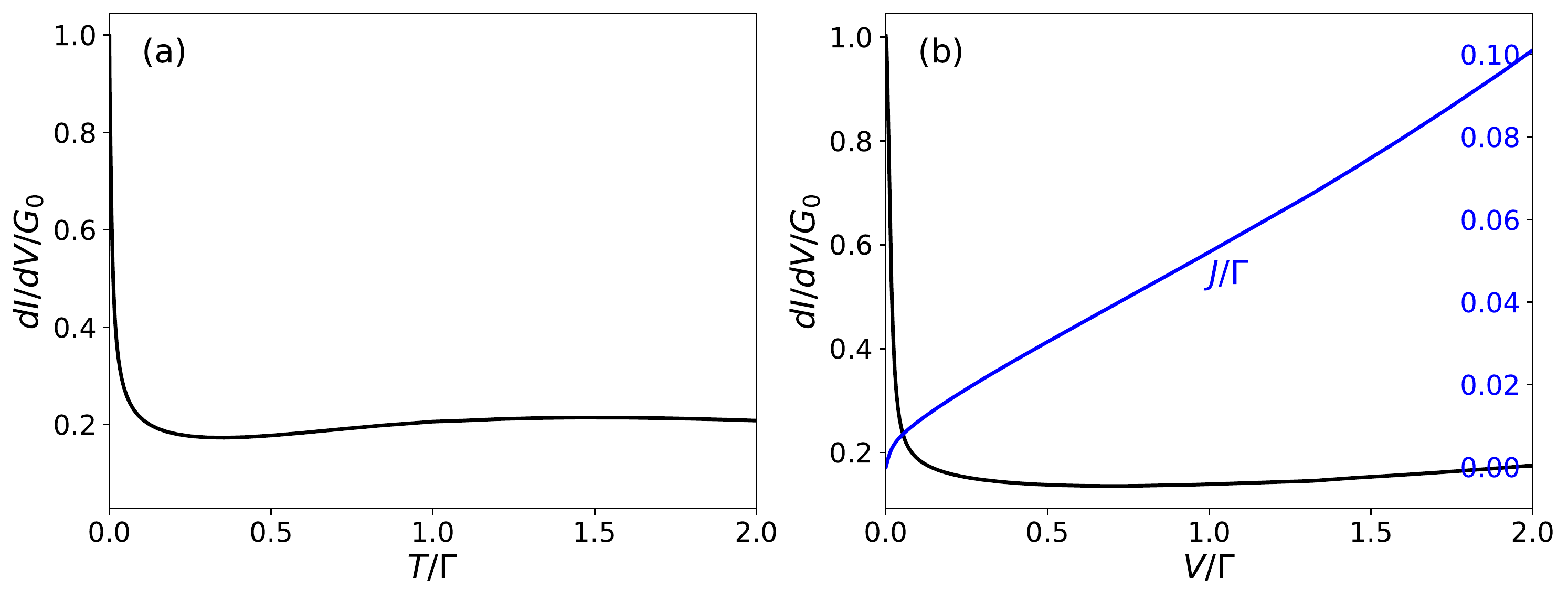}
\caption{Differential conductance (black line) of the SIAM at half-filling as a function of temperature in linear response regime $V=0$ (a) and as a function of voltage far from equilibrium $T=0$ (b). The blue line in the right panel refers to the current.\label{fig: Kondo2CB_conductance}}
\end{figure}

Panel (a) of Fig.~\ref{fig: Kondo2CB_conductance}  plots the differential conductance as a function of temperature at zero bias and  panel (b) as a function of voltage bias at zero temperature.
We denoted $G_0 = 2e^2/h = 1/\pi$  the elementary quantum conductance.
We can observe three transport regimes with the increase of the temperature: Kondo resonance, co-tunneling and sequential tunneling \cite{Pustilnik_2004,bruus2004many}. The zero-bias conductance becomes unity at zero temperature due to the Kondo resonance tunneling \cite{vanderWiel2105}. It remains in the Kondo regime up to the Kondo temperature $T_{K}$ where thermal fluctuations start pushing the electrons away from the Fermi energy which leads to a sharp drop of the conductance. 
When $T_K < T< \Gamma$, the system is driven to the co-tunneling regime where the Kondo peak is effectively suppressed, however, the temperature is not high enough to destroy  coherence in the electron system.
In this regime, the conductance slightly increases since the Hubbard satellite bands contribute to the effective transport energy window. After $T$ crosses $\Gamma$, thermal fluctuations start dominating the system and sequential tunneling plays the major role and the electrons on the QD can be assumed in equilibrium.
For this reason, the conductance will finally decrease to zero in the high-temperature limit.
When we fix the temperature to be zero and increase the voltage $V$, the current (blue line) grows monotonically.
Simultaneously, the differential conductance quickly drops for $V < T_K$ in the Kondo regime but starts growing in the co-tunneling regime when the voltage becomes sufficiently large \cite{PhysRevLett.101.066804,vanderWiel2105}. Notice that the system does not go through the sequential tunneling regime since there are no thermal fluctuations that would destroy coherence of the transport process.

\begin{figure}
	\includegraphics[width=1.0\linewidth]{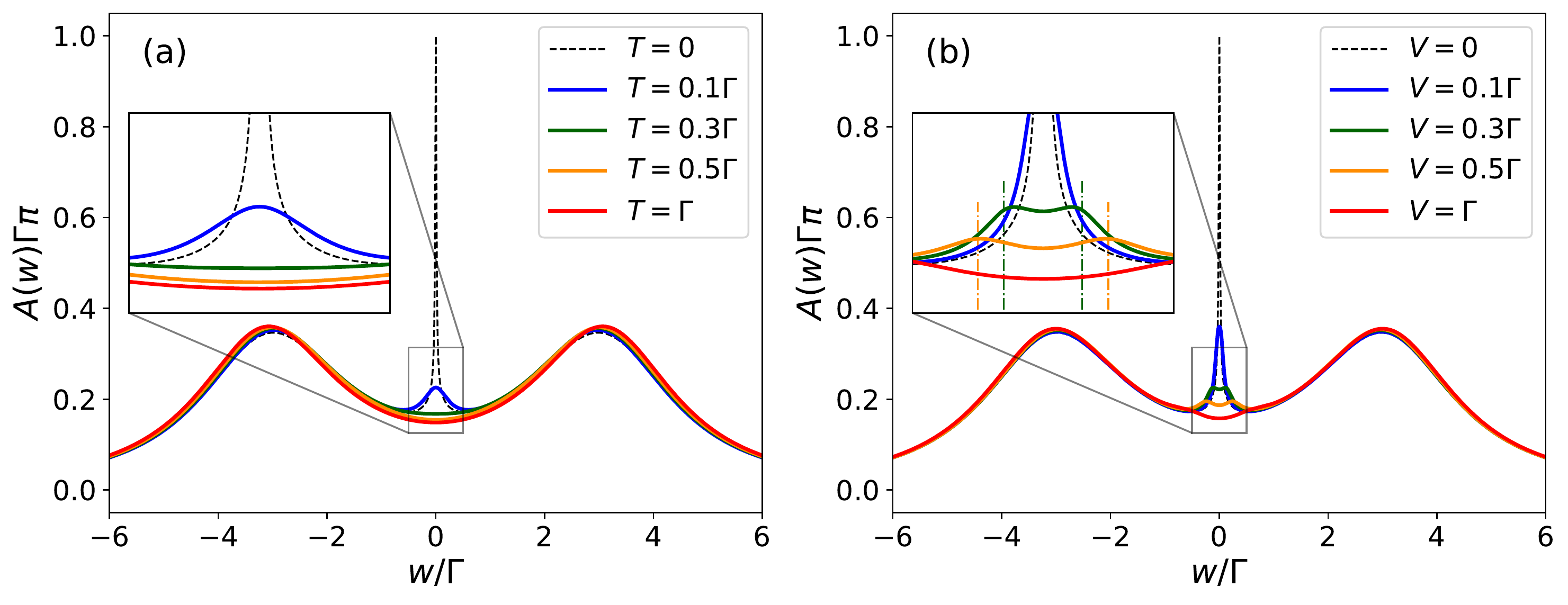}
	\caption{The spectral function of the SIAM  at half filling and for $U=6\Gamma$  at different temperatures and zero bias (a) and different voltage biases and zero temperature (b).
The vertical lines in the inset of panel (b) refer to the local chemical potentials of the leads. \label{fig: Kondo2CB_spectral}}
\end{figure}
\begin{figure}
	\includegraphics[width=1.0\linewidth]{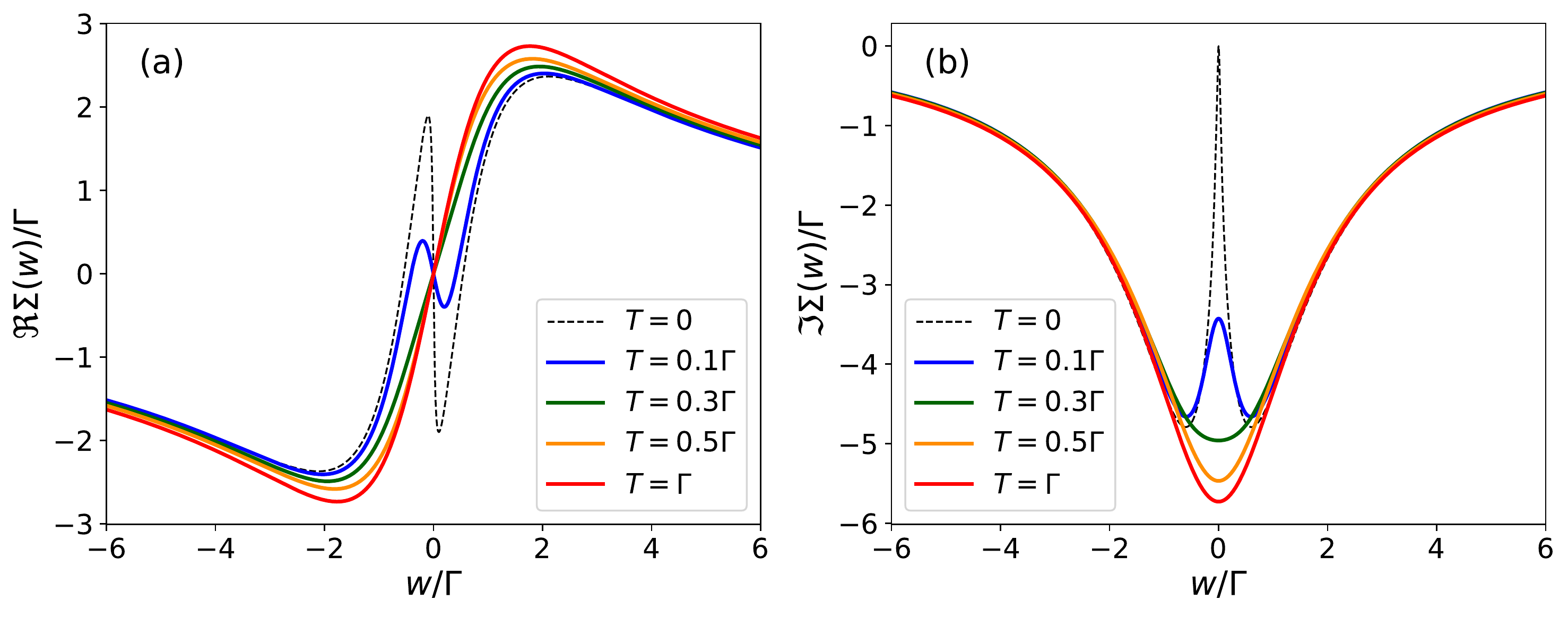}
	\caption{Real and imaginary parts of the spectral self-energy for various temperatures at equilibrium, corresponding to the spectral function plotted in Fig.~\ref{fig: Kondo2CB_spectral} (a). \label{fig: Self-energy}}
\end{figure}

The behavior of the differential conductance in different regimes can be best understood from the spectral function plotted in Fig.~\ref{fig: Kondo2CB_spectral} for various temperatures at zero bias, left pane, and various biases at zero temperature, right pane. The spectral function exhibits a typical three-peak structure with the central narrow quasiparticle peak and the satellite Hubbard bands for $T<T_K$ and $V<T_K$. By comparing Fig.~\ref{fig: Kondo2CB_spectral} (a) and (b), we see that temperature and voltage bias affect similarly the spectral function by broadening the Kondo resonant peak. With the increase of either $T$ or $V$ to $T_K$, the central peak is rapidly suppressed with almost intact satellite bands. The width of the central peak determines a region within which the system behaves as Fermi liquid.
This can be seen from the behavior of the self-energy, plotted for various temperatures in Fig.~\ref{fig: Self-energy}. The bias-dependent self-energy has a similar behavior. The real part of the self-energy has a sharp negative slope and the imaginary part vanishes at the Fermi energy at zero temperature and zero bias. Increasing either temperature or bias the slope of the real part decreases and the imaginary part becomes increasingly negative. Finally, when the slope of the real part of the self-energy turns positive and the local maximum of the imaginary part turns minimum, the Kondo regime is fully destroyed.
Additionally, unlike the temperature-dependent spectral function, the voltage bias further develops local peaks around the chemical potential of each leads \cite{PhysRevLett.70.2601}.  These local peaks are finally destroyed when further increasing the bias, which agrees with the previous experimental results as well as the theoretical studies \cite{PhysRevLett.70.2601,PhysRevLett.101.066804}. The vertical dash lines in the inset of Fig.~\ref{fig: Kondo2CB_spectral} (b) give the positions of local chemical potentials.

%
%

\begin{figure}
	\includegraphics[width=1.0\linewidth]{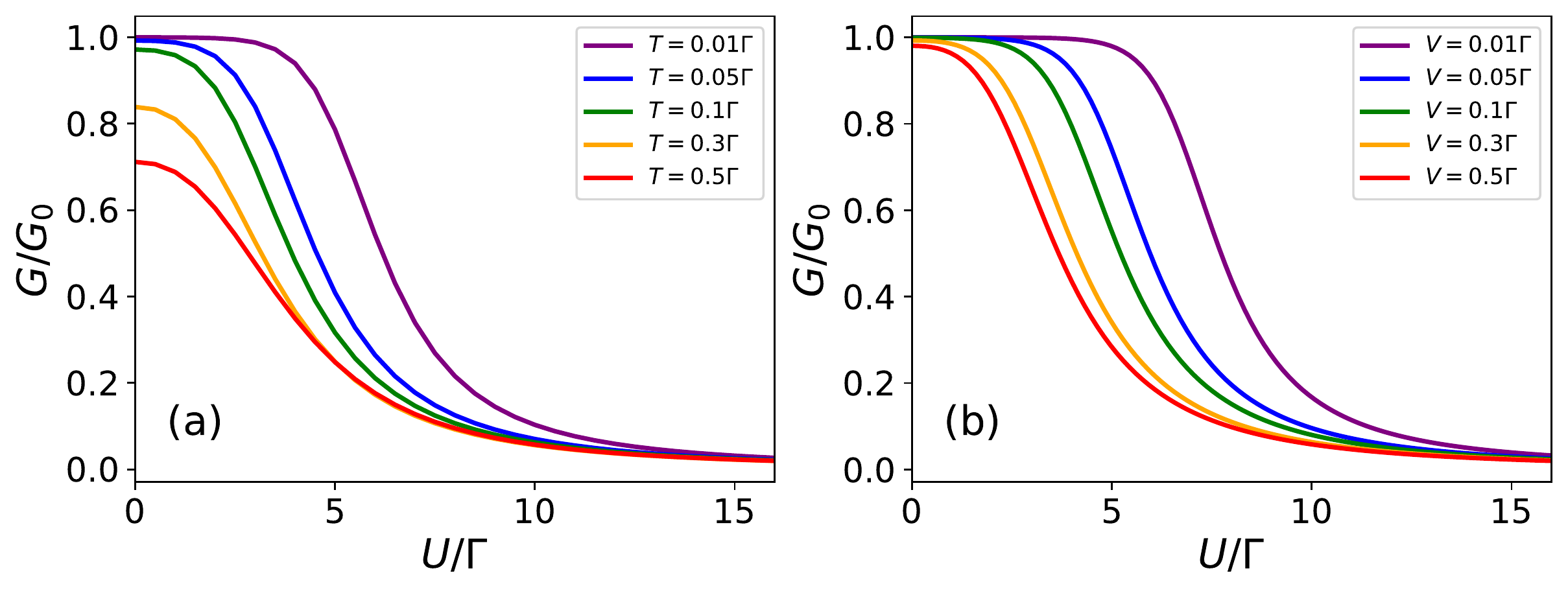}
	\caption{Conductance as a function of the on-site Coulomb repulsion $U$ for various voltage biases $V$.  Linear-response (weak non-equilibrium) regime (left panel) and full non-equilibrium solution (right panel).\label{fig: Udependence}}
\end{figure}

Finally, we also calculated and compared  the conductance as a function of the charging energy $U$ in both the linear-response regime and the fully non-equilibrium solution, plotted in Fig.\ref{fig: Udependence} (a) and (b), respectively.
The conductance in the two extreme cases behaves similarly.
 There is a unity conductance plateau in weak coupling due to the Kondo resonance tunneling \cite{vanderWiel2105}.
 The effective Kondo temperature $T_K$ decreases with the increasing electron repulsion $U$ and once its value becomes comparable with $T$ or $V$, the conductance starts abating with progressive destruction of the Kondo peak. The conductance is significantly suppressed in the extreme limit $U \gg T_K$ due to the Coulomb blockade.

\section{Conclusions\label{sec: conclusions}}

We extended a two-particle semi-analytic approach with the reduced parquet equations and an effective-interaction approximation to an out-of-equilibrium single impurity Anderson model coupled to two biased metallic leads.
The theory was formulated in the critical region of the strong-coupling Kondo limit, capturing the dominant contributions from the spin-flip fluctuations in the instantaneous screened-interaction approximation. It self-consistently determines thermodynamic and spectral quantities.
The reduced parquet equations become analytically solvable and reproduce the logarithmic Kondo scaling in the strong-coupling limit. Numerical solutions are used beyond the Kondo critical regime. We reached a qualitative agreement with experimental and more demanding advanced computational techniques.
Specifically, the hysteresis loop in the current-voltage characteristics, caused by the spurious phase transition weak-coupling approximations, is fully suppressed in the deep Coulomb-blockade regime, $T \gg T_K$ or $V \gg T_K$, due to the renormalization of the effective interaction. We reproduced qualitatively well the temperature dependence of the zero-bias conductance with three stages: Kondo resonant tunneling, $T < T_K$, co-tunneling, $T_K < T < \Gamma$, and sequential tunneling, $T > \Gamma$.
If one keeps temperature zero and turns on the bias, the system will experiencea crossover  from the Kondo resonant regime when $V < T_K$ to a co-tunneling regime when $V > T_K$.
We proved that the biased voltage plays a similar role as temperature in that they both lead to destroying the central Kondo peak when its value crosses the Kondo temperature $T_K$.
Additionally, the applied voltage also tends to develop peaks around the local chemical potentials in low bias.

The theory proved reliable in the electron-hole symmetric case with the qualitatively correct results for the whole range of the model parameters. Some modifications have to be done to extend consistently the present theory to arbitrary filling, away from the Kondo critical region, to keep the compressibility positive.

\begin{acknowledgments}
This work was supported by Grant No.~19-13525S of the Czech Science Foundation.
\end{acknowledgments}

\appendix

\section{Real-time Green Functions\label{sec_app: real time GFs}}

Physical quantities are directly related to real-time Green functions. They can be derived from the Keldysh Green function depending on which branch the real-time temporal arguments they lie.
One defines the following four real-time Green functions \cite{haug2008quantum}
\begin{align}
G_\sigma^t(t,t') &= -i\avg{T_t\{ d_\sigma(t),d_\sigma^\dag(t') \}},\\
G_\sigma^<(t,t') &= +i\avg{d_\sigma^\dag(t') d_\sigma(t)},\\
G_\sigma^>(t,t') &= -i\avg{d_\sigma(t) d_\sigma^\dag(t')},\\
G_\sigma^{\bar{t}}(t,t') &= -i\avg{T_{\bar{t}}\{ d_\sigma(t), d_\sigma^\dag(t') \}},
\end{align}
corresponding to $(t,t')$ sitting on $(-,-)$, $(-,+)$, $(+,-)$ and $(+,+)$ branches, respectively. (We used $-$ for the forward branch and $+$ for the backward branch)
Since the above Green functions are linearly dependent, for further ease of use, we introduce other three linearly independent real-time Green functions
\begin{align}
G_\sigma^r(t,t') &= -i\theta(t-t')\avg{\{ d_\sigma(t), d_\sigma^\dag(t') \}},\\
G_\sigma^a(t,t') &= +i\theta(t'-t)\avg{\{ d_\sigma(t), d_\sigma^\dag(t') \}},\\
G_\sigma^k(t,t') &= -i\avg{[d_\sigma(t), d^\dag_\sigma(t')]}.
\end{align}
In the above formulae we denoted  $\{A,B\} = AB+BA$ and $[A,B] = AB-BA$ the anticommutator and commutator, respectively.
As a result, $G_\sigma^t$, $G_\sigma^<$, $G_\sigma^>$ and $G_\sigma^{\bar{t}}$ can be expressed as a linear combination of $G_\sigma^r$, $G_\sigma^a$ and $G_\sigma^k$ via \cite{PhysRevB.94.045424}
\begin{dmath}\label{eq: linear transform of real-time Green functions}
\begin{bmatrix}
G_\sigma^t & -G_\sigma^<\\
G_\sigma^> & -G_\sigma^{\bar{t}}
\end{bmatrix}
=
\frac{1}{2}
\begin{bmatrix}
G_\sigma^r + G_\sigma^a + G_\sigma^k & G_\sigma^r - G_\sigma^a - G_\sigma^k\\
G_\sigma^r - G_\sigma^a + G_\sigma^k & G_\sigma^r + G_\sigma^a - G_\sigma^k
\end{bmatrix},
\end{dmath}
where we used identities $G_\sigma^> + G_\sigma^< = G^t_\sigma + G^{\bar{t}}_\sigma$ and $G_\sigma^r - G_\sigma^a = G_\sigma^> - G_\sigma^<$.
One can similarly define the bosonic Green functions with the proper change of the quantum statistics. They satisfy the same relations defined above.

\section{Lead Self-energy\label{sec_app: lead self-energy}}

The self-energy of the $s$-lead  can formally be written as
\begin{dmath}\label{eq: lead self-energy}
\Sigma_{s\sigma}^{ld}(z_1,z_2) = \sum_{k} t_{sk} t_{sk}^* G^{cc,0}_{s,kk,\sigma}(z_1,z_2),
\end{dmath}
where
\begin{dmath}\label{eq: decoupled lead Green function}
G^{cc,0}_{s,kk',\sigma}(z_1,z_2) = -i\avg{T_\supset\{ c_{sk\sigma},(z_1), c^\dag_{sk'\sigma}(z_2) \}}_0,
\end{dmath}
is the decoupled c-electron Keldysh Green function. (Subscript `0' implies that the average is taken over the lead Hamiltonian only.)
Since the lead is assumed to be in local equilibrium and in the frequency domain, we have
\begin{dmath}
G^{cc,0,r}_{s,kk,\sigma}(w) = \frac{1}{w-\epsilon_{sk\sigma}+\mu_s + i\eta}.
\end{dmath}
Therefore, the self-energy of the $s$-lead becomes
\begin{dmath}
\Sigma_{s\sigma}^{ld,r}(w) = \frac{1}{\pi} \int_{-\infty}^\infty dx \frac{\Gamma_{s\sigma}(x)}{w-x+i\eta},
\end{dmath}
where $\Gamma_{s\sigma}(x) = \pi \rho_{s\sigma}(x) t_{sk}(x)t^*_{sk}(x)$ is the linewidth function and $\rho_{s\sigma}(x) = \sum_k \delta(x - \epsilon_{sk\sigma} + \mu_s)$ is the spin-resolved density of states of the $s$-lead.

In wide-band limit (WBL), we assume $\Gamma_{s\sigma}(x) = \Gamma_{s\sigma} \theta(|x-\mu_s|<D)$ where $\theta(x)$ is the Heaviside step function. The $s$-lead self-energy then is
\begin{dmath}
\Sigma_{s\sigma}^{ld,r}(w)
= \frac{\Gamma_{s\sigma}}{\pi} \ln\left| \frac{D+(w-\mu_s)}{D-(w-\mu_s)} \right| - i \Gamma_{s\sigma}\theta(|w-\mu_s|<D).
\end{dmath}
We further set $D\rightarrow \infty$, $\Sigma_{s\sigma}^{ld,r}(w) = -i \Gamma_{s\sigma}$.
The lesser and greater $s$-lead self-energies can be obtained by fluctuation dissipation theorem \cite{haug2008quantum}
\begin{dmath}
\Sigma_{s\sigma}^{ld,<}(w) = -2i f_s(w) \Im \Sigma_{s\sigma}^{ld,r}(w),
\end{dmath}
\begin{dmath}
\Sigma_{s\sigma}^{ld,>}(w) = 2i[1-f_s(w)] \Im \Sigma_{s\sigma}^{ld,r}(w),
\end{dmath}
where $f_s(w) = 1/(e^{\beta (w - \mu_s)} + 1)$ is the Fermi-Dirac distribution function.

\section{Derivation of Eq.\eqref{eq: renormalization integral -- steady-state}\label{sec_app: renormalization integral}}

We start from Eq.\eqref{eq: RPE for renormalization integral -- Keldysh contour}.
The complex variable $z$ in $R(z)$ can be chosen either on the forward or backward branch, which is irrelevant to the final result.
By applying the Langreth's rules \cite{haug2008quantum} we obtain 
\begin{dmath}
R^{-}(t) = i\int_{-\infty}^\infty dt_1 dt_2 
\left[
\phi^t(t,t_1)\tilde{\mathcal{K}}^{t}(t_1,t_2)\phi^t(t_2,t) 
- \phi^t(t,t_1)\tilde{\mathcal{K}}^{<}(t_1,t_2)\phi^>(t_2,t) 
- \phi^<(t,t_1)\tilde{\mathcal{K}}^{>}(t_1,t_2)\phi^a(t_2,t) 
- \phi^<(t,t_1)\tilde{\mathcal{K}}^{r}(t_1,t_2)\phi^>(t_2,t)
\right],
\end{dmath}
\begin{dmath}
R^{+}(t) = i\int_{-\infty}^\infty dt_1 dt_2
\left[
\phi^>(t,t_1)\tilde{\mathcal{K}}^{<}(t_1,t_2)\phi^a(t_2,t) 
+ \phi^>(t,t_1)\tilde{\mathcal{K}}^{r}(t_1,t_2)\phi^<(t_2,t) 
- \phi^{\bar{t}}(t,t_1)\tilde{\mathcal{K}}^{>}(t_1,t_2)\phi^<(t_2,t) 
+ \phi^{\bar{t}}(t,t_1)\tilde{\mathcal{K}}^{\bar{t}}(t_1,t_2)\phi^{\bar{t}}(t_2,t)
\right],
\end{dmath}
where  superscript `-'/`+' refers to $t$ lying on the forward/backward branch.
By consider the relations between the real-time Green functions given in Eq.\eqref{eq: linear transform of real-time Green functions}, after averaging $R^-$ and $R^+$ followed by the Fourier transform, one obtains Eq.\eqref{eq: renormalization integral -- steady-state}.

\section{Formulae at Equilibrium\label{sec_app: equilibrium formulae}}

In equilibrium, the above formulae can be simplified by applying the fluctuation-dissipation theorem \cite{haug2008quantum}, which imposes an additional relation between lesser/greater quantities and the corresponding spectral functions.
Specifically, Eq.\eqref{eq: renormalization integral -- steady-state} reduces to
\begin{dmath}
R = \tilde{\Lambda}^2 \mathcal{P} \int_{-\infty}^\infty b(x) \Im \frac{[\phi^r(x)]^3}{1+\tilde{\Lambda}\phi^r(x)} dx,
\end{dmath}
where $\mathcal{P}\int\cdots dx$ refers to the principle-value integral.
Similarly, the even self-energy, Eq.\eqref{eq: even self-energy r/a -- real frequency}, in equilibrium, is given by $\bar{\Sigma}^{int,r}(w) = \frac{Un}{2} + \bar{\Sigma}^{cor,r}(w)$, where
\begin{dmath}
\bar{\Sigma}^{cor,r}(w) =
- \frac{U\tilde{\Lambda}}{2\pi} \sum_\sigma \mathcal{P}\int_{-\infty}^\infty dx \left[ b(x) \Im \frac{\phi^r(x)}{1+\tilde{\Lambda}\phi^r(x)} \bar{G}^r(w-x) - f(x) \frac{\phi^r(x+w)}{1+\tilde{\Lambda}\phi^r(x+w)} \Im \bar{G}^a(-x) \right],
\end{dmath}
and the odd self-energy, Eq.\eqref{eq: odd self-energy r/a -- real frequency}, is unchanged, i.e. $\Delta \Sigma^r(w) = -\frac{m}{2}\tilde{\Lambda}$.
The calculation of the electron-hole bubble is simplified to
\begin{dmath}
\phi^r(w) = -\frac{1}{\pi} \int_{-\infty}^\infty dx f(x) \left[ \bar{G}^a(x-w) \Im \bar{G}^r(x) + \bar{G}^r(x+w) \Im \bar{G}^r(x) \right],
\end{dmath}
and at zero frequency, it reads
\begin{dmath}
\phi^r(0) = -\frac{1}{\pi} \int_{-\infty}^\infty dx f(x) \Im [\bar{G}^r(x) \bar{G}^r(x)].
\end{dmath}

\bibliography{ref}

\end{document}